\documentclass[12pt]{article}
\pdfoutput=1
\usepackage{tikz}
\usetikzlibrary{matrix,arrows,positioning,shapes,snakes,fadings,decorations.pathmorphing,
decorations.pathreplacing,automata,shadows,decorations.markings,patterns}
\usepackage[thicklines]{cancel}
\usetikzlibrary{calc,shapes.callouts,shapes.arrows}
\usepackage{jheppub}
\usepackage{amsmath,amssymb,euscript,array,mathrsfs,appendix,ctable,marvosym}
\usepackage{arydshln}
\usepackage{todonotes}
\usepackage{graphicx}
\usepackage[normalem]{ulem}

\newcommand\blank[1]{#1}
\renewcommand\blank[1]{}
\def\Buildrel#1\over#2\under#3{\mathrel{\mathop{\kern0pt
#2}\limits^{#1}_{#3}}}

\newcommand{\ST}[1]{\vspace{0.4cm}\noindent{\sl #1}\vspace{0.1cm}}

\def\Lax{{\mathscr L}}
\def\CF{{\cal F}}

\def\JJ{\mathscr{J}}

\newcommand{\sech}{\operatorname{sech}}

\newcommand{\Tr}{\operatorname{Tr}}

\def\Bsigma{{\boldsymbol \sigma}}
\def\Blambda{{\boldsymbol \lambda}}
\def\Bv{{\boldsymbol v}}
\def\Bomega{{\boldsymbol \omega}}

\def\BS{{\boldsymbol S}}

\def\BH{{\boldsymbol H}}

\def\SU{\text{SU}}
\def\U{\text{U}}

\newcommand{\Be}{\boldsymbol{e}}
\newcommand{\Balpha}{{\boldsymbol{\alpha}}}

\def\Dbarslash{\,\,{\raise.15ex\hbox{/}\mkern-12mu {\bar D}}}
\def\Dslash{\,{\raise.15ex\hbox{/}\mkern-12mu D}}
\def\delslash{\,{\raise.15ex\hbox{/}\mkern-9mu \partial}}
\def\Aslash{\,{\raise.15ex\hbox{/}\mkern-13mu A}}
\def\delbarslash{\,\,{\raise.15ex\hbox{/}\mkern-9mu {\bar\partial}}}

\def\ket#1{| #1\rangle}

\def\dint{\displaystyle\int}

\newcommand{\MAT}[1]{\begin{pmatrix} #1\end{pmatrix}}
\newcommand{\ARR}[1]{\begin{matrix} #1\end{matrix}}
\newcommand{\EQ}[1]{\begin{equation}\begin{split} #1
\end{split}\end{equation}}

\title{Quantum Inverse Scattering and the Lambda Deformed Principal Chiral Model}
\author{Calan Appadu, Timothy J. Hollowood and Dafydd Price}
\affiliation{Department of Physics, Swansea University, Swansea, SA2 8PP, U.K.}

\emailAdd{t.hollowood@swansea.ac.uk}

\abstract{The lambda model is a one parameter deformation of the principal chiral model that arises when regularizing the non-compactness of a non-abelian T dual in string theory. It is a current-current deformation of a WZW model that is known to be integrable at the classical and quantum level. The standard techniques of the quantum inverse scattering method cannot be applied because the Poisson bracket is non ultra-local. Inspired by an approach of Faddeev and Reshetikhin, we show that in this class of models, there is a way to deform the symplectic structure of the theory leading to a much simpler theory that is ultra-local and can be quantized on the lattice whilst preserving integrability. This lattice theory takes the form of a generalized spin chain that can be solved by standard algebraic Bethe Ansatz techniques. We then argue that the IR limit of the lattice theory lies in the universality class of the lambda model implying that the spin chain provides a way to apply the quantum inverse scattering method to this non ultra-local theory. This points to a way of applying the same ideas to other lambda models and potentially the string world-sheet theory in the gauge-gravity correspondence.
}

\notoc
\begin{document}

\maketitle

\newpage

\pgfdeclarelayer{top layer}
\pgfdeclarelayer{foreground layer}
\pgfdeclarelayer{background layer}
\pgfsetlayers{background layer,main,foreground layer,top layer}

\section{Introduction}

It is remarkable that the gauge-gravity correspondence has deep within it an integrable structure. The world sheet theory of the string in AdS$_5\times S^5$ is an integrable QFT and the spectral problem for operators in ${\cal N}{=}4$ gauge theory involves a discrete integrable system (see \cite{Beisert:2010jr,Arutyunov:2009ga} for reviews). 

The solution of an integrable QFT in $1{+}1$ dimensions leads to the idea of the Quantum Inverse Scattering Method (QISM).\footnote{This method was developed in many works. Fortunately there exists an excellent review \cite{Faddeev:1996iy} and a book \cite{KorepinBook} which both have many original references.} The essence of the QISM is to define a version of the QFT on a spatial lattice in such a way that integrability is explicitly maintained. The quantum lattice model provides a UV regularized version of the QFT consistent with integrability. The lattice model generally takes the form of a generalized Heisenberg spin chain in its anti-ferromagnetic regime.\footnote{This latter fact is important: relativistic QFTs have non-trivial ground states with spatial entanglement that is not compatible with the ferromagnetic ground state.} The spectrum of the theory can then be solved via a version of the Bethe Ansatz---the so-called Algebraic Bethe Ansatz (ABA) which is the engine of the QISM---and then a continuum limit can be taken with particle states of the QFT being identified with excitations over the non-trivial ground state. Then the S-matrix of the excitations can be extracted and, in principle, the correlation functions can be found \cite{KorepinBook}.

In the string theory context, the goal would be to apply the QISM to the world sheet theory. However, the world sheet theory, being a kind of sigma model suffers from a fundamental problem that thwarts a direct application of the QISM: the classical theory is non ultra-local. This means that the fundamental Poisson brackets of the currents in the theory have a central term that involves the derivative of a delta function $\delta'(x-y)$. There is, as yet, no completely successful way of overcoming this problem and applying the QISM (see \cite{Freidel:1991jx,Freidel:1991jv,Kundu:2002et,Melikyan:2016gkd} for some attempts at a general approach and \cite{Vicedo:2017cge,Lacroix:2017isl} for some recent new ideas). There have been attempts to overcome these difficulties by applying QISM to related models and then taking particular limits to recover the sigma model \cite{Polyakov:1983tt,Faddeev:1985qu}. It is these approaches that inspire the current work.  

In summary, in this paper we:
\begin{enumerate} 
\item Define a deformation of the classical lambda models that is ultra local (this is a version of the {\it alleviation\/} procedure of \cite{Delduc:2012qb,Delduc:2012vq} which is inspired in turn by the original work of Faddeev and Reshetikhin \cite{Faddeev:1985qu}).
\item Show that the ultra local version is amenable to the QISM and show that the underlying regularized theory is a kind of generalized Heisenberg spin chain that lies in the class of regularized theories defined on a null lattice in space time.  
\item Argue that in the continuum limit that the IR excitations around the ground state have a relativistic dispersion relation and a factorizable S-matrix that was known previously to describe the perturbed WZW model.
\item Take the S-matrix of the excitations and use it to calculate the free energy density in the presence of a background charge. The same quantity can also be calculated via perturbation theory from the lambda model and also directly from the spin chain. We find precise agreement for the three methods of calculation.
\item Argue that the QISM techniques can be extended to cover all the bosonic lambda models providing some encouragement that ultimately the QISM can be applied to the string world sheet in the lambda deformed $\text{AdS}_5\times S^5$ background.
\end{enumerate} 

Lying behind this work is the question of how to apply the QISM to the string world sheet sigma model? Rather than tackling this head on, recently various integrable deformation of the theory of world sheet of the string in AdS$_5{\times}S^5$ have been investigated (see the selection of papers \cite{Arutyunov:2012zt,Arutyunov:2012ai,vanTongeren:2013gva,Hollowood:2014qma,Sfetsos:2014cea,Demulder:2015lva,Hoare:2014pna,Hoare:2015gda,Sfetsos:2015nya,Hollowood:2014rla,Hollowood:2015dpa,Vicedo:2015pna,Borsato:2016zcf,Klimcik:2016rov,Chervonyi:2016ajp,Borsato:2016ose,Chervonyi:2016bfl,Sfetsos:2014lla,Georgiou:2016zyo} which consider the kinds of deformation that are relevant to the present work). It seems that one class of these deformations, the lambda deformations, lead to a consistent world sheet theory at the quantum level \cite{Hollowood:2014qma}. The lambda deformations were formulated some time ago---the name being much more recent---as deformations of the Poisson structure of a sigma model that at the classical level continuously deforms it into a WZW model with a current-current perturbation \cite{Balog:1993es,Evans:1994hi}. 

More recently the lambda deformations appeared in the context of non-abelian T duality in string theory. Unlike its abelian cousin, non-abelian T duality is, strictly speaking, not a symmetry of string theory, but rather can be viewed as a way of generating new string backgrounds \cite{delaOssa:1992vci,Alvarez:1993qi,Giveon:1993ai,Alvarez:1994np,Sfetsos:2010uq}. These new backgrounds often have some kind of pathology in the form of singularities and/or non compactness. However, there has been some progress in making sense of these apparently pathological backgrounds. For instance, one can suitable deform the world sheet theory \cite{Sfetsos:2013wia}. Alternatively the singular non-abelian T dual geometry can be  understood as the Penrose limit of a more consistent geometry \cite{Lozano:2016kum,Lozano:2017ole}.

Let us describe the world sheet deformation approach. Suppose the sigma model has a non-abelian symmetry $F$. The non-abelian T dual is defined by gauging the $F$ symmetry and then adding a Lagrange multiplier field in the adjoint of the Lie algebra $\mathfrak f$ in order to enforce the flatness of the gauge field:
\EQ{
\begin{tikzpicture}[scale=0.8] 
\node at (0,0) (a1) {$S_\sigma[f]$};
\node at (7,0) (a2) {$S_{\text{g-}\sigma}[f,A_\mu]-\dint d^2x\,\Tr(\nu F_{+-})$};
\node at (4,-3) (a3) {$-\dfrac{\kappa^2}{4\pi}\dint d^2x\,\Tr(A_\mu A^\mu)-\dint d^2x\,\Tr(\nu F_{+-})$};
\draw[->,thick] (a1) -- (a2);
\node at (3.5,-1.2) {\small gauge fixing $f=1$};
\draw[->,thick] (6,-0.5) -- (6,-2);
\begin{scope}[yshift=-0.5cm]
\draw[-,thick] (4,-3.5) -- (4,-4);
\draw[<->,thick] (2,-5.5) -- (2,-4) -- (6,-4) -- (6,-5.5);
\node at (0,-4.2) {\small integrate out $\nu$};
\node at (0,-4.8) {\small $A_\mu=f^{-1}\partial_\mu f$};
\node at (8,-4.5) {\small integrate out $A_\mu$};
\node at (2,-6) (a4) {$S_\sigma[f]$};
\node at (6,-6) (a5) {$S_\text{NATD}[\nu]$};
\end{scope}
\end{tikzpicture}
\label{llk}
}
The gauge symmetry can then be fixed by setting $f=1$.
Integrating out $\nu$, enforces the flatness of the gauge field which implies $A_\mu=f^{-1}\partial_\mu f$ and returns us to the original sigma model while integrating out the gauge field gives us the non-abelian T-dual of the sigma model where now $\nu$ is the fundamental field.

But $\nu$ being Lie algebra valued is a non-compact field even when the group $F$ is compact. This can complicate the geometrical interpretation of the non-abelian T dual. In order to proceed, a kind of regularization procedure can be followed that  replaces the Lagrange multiplier field by an $F$-group valued field $\CF$ and the coupling in \eqref{llk} by the gauged WZW action  \cite{Sfetsos:2013wia}:
\EQ{
\int d^2x\,\Tr(\nu F_{+-})\longrightarrow kS_\text{g-WZW}[\CF,A_\mu]\ .
} 
We remark that the new action is the gauged WZW model where the full vector $F$ symmetry is gauged---we denote this as $F/F_V$---and it has a new coupling constant, the level $k$. The intuition---as we will see somewhat na\"\i ve---is that in the $k\to
\infty$ limit, the WZW model is effectively classical, so fluctuations are suppressed and expanding the group element $\CF$ around the identity as $\CF=\exp[\nu/\sqrt k]$, one recovers \eqref{llk}. 

The lambda model is then obtained from the extended action
\EQ{
S[f,\CF,A_\mu]=S_{\text{g-}\sigma}[f,A_\mu]+kS_\text{g-WZW}[\CF,A_\mu]\ ,
\label{ppo}
}
by gauge fixing the $F$ gauge symmetry, i.e.~by choosing a gauge slice $f=1$ (when $F$ acts freely). In the gauge fixed theory the gauge field itself $A_\mu$ becomes an auxiliary field that can be integrated out. 

What makes the lambda deformation particularly interesting is that if the original sigma model is integrable, then the associated lambda model is also integrable. Note that this applies strictly to the integrable bosonic theories. For the cases appropriate to the superstring, the group is a super group and the issue of preserving integrability is more subtle and in particular a na\"\i ve use of \eqref{ppo} does not lead to an integrable theory \cite{Hollowood:2014qma}. 

There are two classes of integrable sigma models for which it is interesting to apply the lambda deformation, the Principal Chiral Models (PCMs) and the symmetric space sigma models. We will only consider the PCMs in the present work (although we make some comments and conjectures on the symmetric space theories in the final section). These theories have a field  $f$ valued in the group $F$, with an action
\EQ{
S_\sigma[f]=-\frac{\kappa^2}{4\pi}\int d^2x\,\Tr(f^{-1}\partial_\mu f\,f^{-1}\partial^\mu f)\ .
}
The theory enjoys a $F_L\times F_R$ global symmetry $f\to U_LfU_R^{-1}$ and therefore there are two distinct kinds of lambda deformation depending upon whether ones chooses to apply the deformation only on the symmetry subgroup $F_L$ (or equivalently $F_R$) or the full $F_L\times F_R$. Note that only these two choices leads to  integrable lambda theories.

Turning to the PCM and focussing on the lambda associated to the symmetry group $F_L$, if we follow the procedure above, then gauge fixing the extended theory \eqref{ppo} gives rise to the lambda model 
\EQ{
S_\lambda[\CF,A_\mu]=kS_\text{g-WZW}[\CF,A_\mu]-\frac{\kappa^2}{4\pi}\int d^2x\,\Tr(A_\mu A^\mu)\ .
\label{tyy}
}
The second term here is just the gauged sigma model action with $f=1$. 

This second term in \eqref{tyy} has dramatic effects: the equations of motion of $A_\mu$ in the gauged WZW model change from first class constraints into second class constraints indicating that strictly-speaking $A_\mu$ is not a gauge field but just a Gaussian auxiliary field that can be integrated out. 

The monicker ``lambda model" derives from the fact that one introduces the coupling
\EQ{
\lambda=\frac k{k+\kappa^2}\ .
}
In the limit, $\lambda\to0$, the field $A_\mu$ freezes out and the theory becomes the (non-gauged) WZW model for the group $F$ with level $k$. At the classical level, $\lambda$ parameterizes a family of integrable classical field theories. 

At the quantum level, conformal invariance is broken and $\lambda$ runs, increasing from from $0$ into the IR: it is a marginally relevant coupling. There is an exact expression for the beta function to leading order in $1/k$ \cite{Tseytlin:1993hm,Sfetsos:2014jfa,Appadu:2015nfa}:\footnote{$c_2(F)$ is the quadratic Casimir of the adjoint representation, the dual Coxeter number, of $F$. For $\SU(N)$, $c_2(F)=N$.}
\EQ{
\mu\frac{d\lambda}{d\mu}=-\frac{2c_2(F)}{k}\Big(\frac\lambda{1+\lambda}\Big)^2\ ,
\label{bft}
}
where $c_2(F)$ is the dual Coxeter number of $F$.
The non-gauged WZW model describes the UV limit. For small $\lambda$, one finds that the action (after integrating out $A_\mu$) takes the form of a current-current deformation of the UV WZW model
\EQ{
S_\lambda[\CF]=kS_\text{WZW}[\CF]+\frac{4\pi\lambda}k \int d^2x\,\Tr\big(\hat\JJ_+\hat\JJ_-\big)+{\cal O}(\lambda^2)\ ,
\label{bxx}
}
where $\hat\JJ_\pm$ are the usual affine currents of the WZW model:
\EQ{
\hat\JJ_+=-\frac k{2\pi}\CF^{-1}\partial_+\CF\ ,\qquad\hat\JJ_-=\frac k{2\pi}\partial_-\CF\CF^{-1}\ .
}
In this limit, the theory can be described as a marginally relevant deformation of the WZW CFT thats leads to a massive QFT  that is known to preserve integrability \cite{Ahn:1990gn,Bernard:1990jw,Evans:1994hi}. This is entirely consistent with the classical analysis that shows that integrability can be maintained for any $\lambda$. In the quantum theory, $\lambda$ will transmute into the mass gap of the massive QFT.

Returning to the QISM, we can ask whether the lambda models are in better shape than the PCM as regards the problem of  non ultra-locality? The answer is no, they are still non ultra-local; for instance this is clear because the affine currents of the theory obey a classical version of the Kac-Moody algebra including the central term proportional to $\delta'(x-y)$. 

So non ultra-locality is still present. We take as inspiration, however, that at the level of the Poisson brackets which depend on the two parameters $(k,\lambda)$, the classical theory admits a subtle limit that involves taking $k\to0$ and $\lambda\to0$ keeping the ratio 
\EQ{
\nu=\frac k{4\pi\lambda}
\label{rte}
}
fixed. This limit cannot literally be taken at the level of the quantum lambda model because that must have integer $k$ coming from the consistency of the functional integral in the presence of the WZ term in the action. However, the limit makes perfect sense at the level of the classical Hamiltonian structure. It turns out that this limit is an example of the ``alleviation procedure" of \cite{Delduc:2012qb,Delduc:2012vq}. In the limit, the theory becomes ultra-local and the QISM becomes available. In fact, for the case $F=\SU(2)$ the limiting theory is precisely that considered by Faddeev and Reshetikhin \cite{Faddeev:1985qu} in their attempt to apply QISM to the PCM. We will call the limiting theory for a general group, the Linear Chiral Model (LCM). The LCM model has as a parameter $\nu$, identified as \eqref{rte} above  but also the values of certain centres of the Poisson algebra. These centres can be identified with Casimirs of the associated Lie algebra $\mathfrak f$  and in the quantum theory they correspond to fixing a particular representation of the group $F$. For the application to the lambda model the appropriate representation, for the case $F=\SU(N)$, is the rank $k$ symmetric representation and this determines the type of local spin of the spin chain that results from the QISM (for $\SU(2)$ the spin is $k/2$).

\begin{figure}
\begin{center}
\begin{tikzpicture}[scale=0.8,fill=black!10] 
\draw[fill] (-2.5,-2.5) -- (-2.5,3.5) -- (2.5,2.5) -- (2.5,-3.5) --(-2.5,-2.5);
\begin{scope}[very thick,decoration={markings,mark=at position 0.5 with {\arrow{>}}}] 
    \draw[postaction={decorate}] plot[smooth] coordinates {(-1,3) (-0.5,1.7) (0,0)};    
   \draw[postaction={decorate}]  (0,0) -- (-5,-2);
     \draw[postaction={decorate}] plot[smooth] coordinates {(-1,2) (-0.9,0.1) (-5,-1.95)};     
      \draw[postaction={decorate}] plot[smooth] coordinates {(-1.5,1.75) (-1.7,-0.1) (-5,-1.9)};      
     \draw[postaction={decorate}] (-2.2,1.4) to[out=280,in=25] (-5,-1.85);   
   \draw[postaction={decorate}] (-3,1) to[out=280,in=25] (-5,-1.8);   
     \end{scope}
\filldraw[black] (-0.68,2.19) circle (2pt);
\filldraw[black] (0,0) circle (2pt);
\filldraw[black] (-1.5,1.75) circle (2pt);
\filldraw[black] (-3,1) circle (2pt);
\filldraw[black] (-2.2,1.4) circle (2pt);
\filldraw[black] (-1,2) circle (2pt);
\draw[densely dashed] (-4,0.5) -- (-3,1) -- (-2,1.5) -- (-1,2) -- (-0.7,2.2);
\node at (-5.5,-1.9) (n6) {IR};
\draw[->] (-5,0.5) -- (-4,1);
\node[rotate=27] at (-4.8,0.9) (n1) {$\nu$ increasing};
\node at (-3.5,2.5) (n2) {LCM$(\nu,k)$ model};
\node at (1,2.2) (n3) {$\nu{=}\infty$};
\draw[->] (n3) -- (-0.5,2.2);
\draw[->] (n2) -- (-2.5,1.4);
\node at (1,-1) (p1) {WZW$_k$};
\draw[->] (p1) -- (0.2,-0.2);
\node at (0,-2) (p2) {lambda model};
\draw[->] (p2) -- (-1.2,-0.6);
\node at (4.5,2) (n3) {critical surface};
\draw[->] (n3) -- (2,1.8);
\node at (4.7,0.5) (n4) {crossover};
\draw[->] (n4) -- (-0.5,0.5);
\node at (-3.5,-4) (n5) {massive theory};
\draw[->] (n5) -- (-4.6,-1.9);
\end{tikzpicture}
\caption{\footnotesize  The picture of renormalization group flows that we want to establish. The LCM is associated to a rank $k$ symmetric representation of $F$ identified with the level of the WZW model of the UV fixed point.}
\label{f1} 
\end{center}
\end{figure}
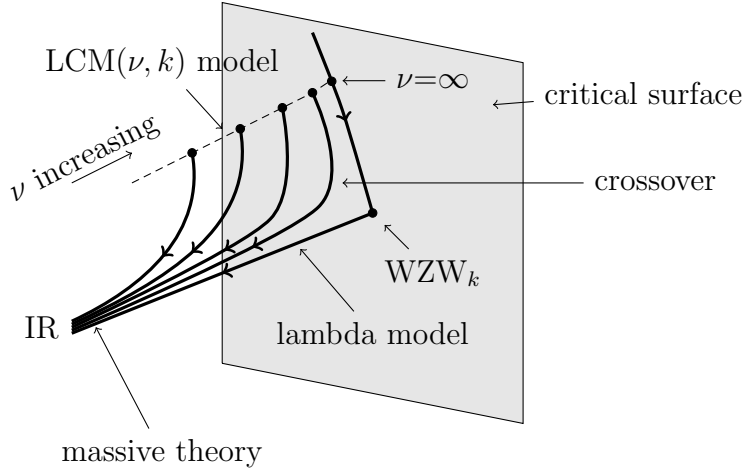

The above construction of a new theory, the LCM model, which can now be tacked by QISM is mildly diverting by itself, but  can it teach us anything about the lambda model and ultimately in the $k\to\infty$ limit the PCM? We will provide strong evidence that the LCM theory has a continuum limit that describes a massive relativistic QFT lying in the universality class of the lambda model with a level $k$. The continuum limit involves taking $\nu\to\infty$ and lattice spacing $\Delta\to0$ with the combination
\EQ{
\frac1\Delta\exp\big[-2\pi \nu/c_2(F)\big]=\text{fixed}\ .
}
In fact, one can easily check that this is precisely consistent with the beta function in \eqref{bft} in the UV limit $\lambda\to0$, with the spatial cut off $\Delta$ identified with $\mu^{-1}$. This gives us some immediate preliminary evidence that the LCM model does indeed describe the lambda model in the IR. However, the evidence is much stronger than this. The spectrum of excitations of spin chain and their S-matrix precisely matches those of the lambda model that were previously deduced by using conformal perturbation theory techniques \cite{Ahn:1990gn} and Thermodynamic Bethe Ansatz (TBA) techniques \cite{Evans:1994hi}. The picture of renormalization group flows that we are proposing is shown in fig.~\ref{f1}.

Another aspect of the RG flow is that by focussing on length scales much shorter than the correlation length but much greater than the cut off, we can tune into the crossover in the neighbourhood of the WZW fixed point. 

The papers is organized as follows: in section \ref{s2} we introduce the lambda model describing its conserved charges and Poisson brackets pointing out the non ultra-locality. In section \ref{s3}, we show how a simpler ultra-local theory---the LCM---can be defined by taking a suitable limit. In section \ref{s4}, we show how the LCM can be discretized and quantized on a null lattice in 2d Minkowski space following the formalism of Destri and de Vega \cite{Destri:1987ze,Destri:1987hc,Destri:1987ug} and Faddeev and Reshetikhin \cite{Faddeev:1985qu,Faddeev:1996iy}. This leads to a spin chain which can be tackled, as we show in section \ref{s5}, by the algebraic Bethe Ansatz. In this section, we discuss the ground state and excitations and their S-matrix. In section \ref{s6}, we then provide a strong test of the hypothesis summarized in fig.~\ref{f1}, by showing that when the theory is coupled to a conserved charge, the shift in free energy calculated from (i) the S-matrix via the thermodynamic Bethe Ansatz; (ii) the spin chain; and (iii) the WZW Lagrangian via perturbation theory all agree. In section \ref{s7} we draw some conclusions and suggest how the spin chain construction of the QISM can be generalized to the integrable symmetric space lambda models.

\section{The classical lambda models and non ultra-locality}\label{s2}

In this section, we analyse the lambda models at the classical level focussing on their integrability, Poisson structure and highlighting the issue of non ultra-locality \cite{Hollowood:2014rla}.

The action for the theory takes the form \eqref{tyy}.
The equation of motion of the group field $\CF$ is not affected by the deforming $A_+A_-$ term and can be written either as\footnote{We take 2d metric $\eta_{\mu\nu}=\text{diag}(1,-1)$. We often use the null coordinates $x^\pm=t\pm x$ and for vectors we have $A^\pm=A^0\pm A^1$ and $A_\pm=(A_0\pm A_1)/2$ so that the invariant $A_\mu B^\mu=2(A_+B_-+A_-B_+)$.}\EQ{
\big[\partial_++\CF ^{-1}\partial_+\CF +\CF ^{-1}A_+\CF ,\partial_-+A _-\big]=0\ ,
\label{sd2}
}
or, equivalently, by conjugating with $\CF$, as
\EQ{
\big[\partial_++A _+,\partial_--\partial_-\CF \CF ^{-1}+\CF  A_-\CF ^{-1}\big]=0\ .
\label{sd3}
}
Since the action for auxiliary field $A_\mu$ has no derivatives, its equations of motion take the form of the constraints
\EQ{
\CF ^{-1}\partial_+\CF +\CF ^{-1}A_+\CF&=\frac1\lambda A_+\ ,\\
-\partial_-\CF \CF ^{-1}+\CF A_-\CF ^{-1}&=\frac1\lambda A_-\ .
\label{con1}
}
In the constrained Hamiltonian formalism of Dirac these constraints are second class and so can be imposed strongly on the phase space (for details of this see \cite{Hollowood:2014rla}). If we use these constraints, then the equations of motion for the group field \eqref{sd2} or \eqref{sd3} can be written solely in terms of the auxiliary field $A_\mu$,
\EQ{
&-\partial_-A _++\lambda \partial_+A _-+[A_+,A_-]=0\ ,\\
&-\lambda \partial_-A _++\partial_+A _-+[A_+,A_-]=0\ ,
\label{pls}
}
from which we find (for $\lambda\neq1$)
\EQ{
\partial_\mp A_\pm=\pm\frac1{1+\lambda}[A_+,A_-]\ .
\label{ger}
}

The auxiliary field $A_\mu$ can be related to the usual Kac-Moody currents of the $F/F_V$ gauged WZW model,
\EQ{
\JJ_+&=-\frac k{2\pi}\big(\CF ^{-1}\partial_+\CF +\CF ^{-1}A_+\CF -A _-\big)\ ,\\
\JJ_-&=\frac k{2\pi}\big(\partial_-\CF \CF ^{-1}-\CF A_-\CF ^{-1}+A _+\big)\ .
\label{ksq2}
}
Given the constraints \eqref{con1}, we have
\EQ{
\JJ_\pm=-\frac k{2\pi}\Big(\frac1{\lambda}A_\pm-A _\mp\Big)\ .
\label{sd1}
}
We can invert these relations, assuming $\lambda\neq1$, to express the auxiliary field $A_\mu$ in terms of the Kac-Moody currents,
\EQ{
A_\pm=-\frac{2\pi\lambda }{k(1-\lambda^2)}\big(\JJ_\pm+\lambda \JJ_\mp\big)\ .
\label{uu8}
}

\ST{Conserved charges} 

The  equations of motion \eqref{pls} can be written directly as a Lax equation for a connection $\Lax_\mu(z)$
\EQ{
\Lax_\pm(z)=\frac{2\nu  }{\nu\pm z}\frac1{1+\lambda}A_\pm\ ,
\label{lcc}
}
where $z$ is the spectral parameter, a free variable whose existence underlies the integrability of the theory, and $\nu$ is a free constant. The Lax equation is the flatness condition for $\Lax_\mu(z)$ for arbitrary $z$:
\EQ{
[\partial_++\Lax_+(z),\partial_-+\Lax_-(z)]=0\ .
\label{leq}
}
In particular, the form of the equations of motion \eqref{ger} follow from the residues at the poles $z=\pm\nu $.

The fact that the equations of motion can be written in Lax form is the key to unlocking the integrability of the classical theory.
Since $\Lax_\mu(z)$ is a flat connection the spectrum of the monodromy matrix
\EQ{
T(z)=\text{Pexp}\Big[-\int_{-L}^L dx\,\Lax(x;z)\Big]\ ,
}
with $\Lax\equiv\Lax_x$,
assuming periodic boundary conditions,\footnote{Or in an infinite space, with suitable fall off of fields at $x=\pm\infty$.} is conserved in time. The fact that $T(z)$ depends on the spectral parameter $z$ means that it is a generating function for an infinite set of conserved charges.

It is a standard feature in integrable system \cite{BBT}, that sets of conserved charges can be constructed around each of the poles $z=\pm\nu $ of the Lax connection. The idea is to construct gauge transformations
\EQ{
\Lax^{(\pm)}_\mu(z)=V^{(\pm)}(z)^{-1}\Lax_\mu(z) V^{(\pm)}(z)+V^{(\pm)}(z)^{-1}\partial_\mu V^{(\pm)}(z)\ ,
}
regular at the poles,
\EQ{
V^{(\pm)}(z)=\sum_{n=0}^\infty (z\pm\nu )^nV^{(\pm)}_n\ ,
}
that abelianize the Lax connection. These gauge transformations can be found order by order in $z\pm\nu $. Concretely this means that $\Lax^{(\pm)}_\mu$ lies in a Cartan subalgebra of $\mathfrak f$ and so
\EQ{
\partial_+\Lax^{(\pm)}_-(z)-\partial_-\Lax^{(\pm)}_+(z)=0\ .
}
It follows that the coefficients in the expansions in $\nu\pm z$ are conserved currents:
\EQ{
\Lax_\mu(z)=\sum_{n=-1}^\infty (z\pm\nu )^n\Lax^{(\pm)}_{n,\mu}\ ,
}
with
\EQ{
\partial_\mu \big(\epsilon^{\mu\nu}\Lax^{(\pm)}_{n,\nu}\big)=0\ .
}
The associated charges, valued in the Cartan subalgebra, are given by
\EQ{
I^{(\pm)}_n=\int_{-L}^L dx\,\Lax^{(\pm)}_n\ ,\qquad \Lax_n^{(\pm)}\equiv\Lax^{(\pm)}_{n,x}\ .
}
These charges determine the spectrum of the monodromy matrix via
\EQ{
T(z)=V^{(\pm)}(L)^{-1}\exp\Big[-\sum_{n=-1}^\infty I^{(\pm)}_n(z\pm\nu )^n\Big]V^{(\pm)}(L)\ .
}
where we have imposed periodicity on the gauge transformation 

However, there are other conserved charges that can be constructed from the Lax connection that are also associated to local conserved currents. These follow from the observation that since
$\Lax_{-1,\mp}^{(\pm)}=0$, we have the chiral conservation equations
\EQ{
\partial_\mp \Lax_{-1,\pm}^{(\pm)}=0\ .
}
This means that products of traces of powers of 
\EQ{
\Lax_{-1,\pm}^{(\pm)}=\frac{2\nu  }{1+\lambda}V^{(\pm)-1}_0A_\pm V_0^{(\pm)}
}
also yield conserved charges. The simplest set are associated to single traces of the form
\EQ{
Q^{(\pm)}_n=\int_{-L}^L dx\,\Tr\big[\big(A_\pm\big)^n\big]
}
This second set of conserved charges includes the Hamiltonian of the theory
\EQ{
{\cal H}&=-\frac{k(1-\lambda^2)}{4\pi}\big(Q_2^{(+)}+Q_2^{(-)}\big)\\
&=-\frac{\pi}{k(1-\lambda^2)}\int_{-L}^L dx\,\Tr\Big[(1+\lambda^2)(\JJ_+\JJ_+ +\JJ_-\JJ_-)+4\lambda \JJ_+\JJ_-\Big]\ .
}
In fact the non-vanishing components of the energy-momentum tensor are
\EQ{
T_{\pm\pm}=-\frac{k(1-\lambda^2)}{4\pi}\Tr\big[(A_\pm)^2\big]\ .
}
So the momentum is equal to
\EQ{
{\cal P}=\frac{k(1-\lambda^2)}{4\pi}\big(Q_2^{(+)}-Q_2^{(-)}\big)\ .
}

The theory also has a set of non-local conserved charges that are associated to the expansion of the monodromy matrix around $z=\infty$ but we shall not need them here.

\ST{Poisson brackets}

In the Hamiltonian formalism, the reduced phase space, after imposing the constraints \eqref{con1} strongly on phase space, is parameterized by either $A_\mu$ or $\JJ_\mu$. The Poisson brackets of $\JJ_\mu$ on the reduced phase space are equal to those on the original phase space and these take the form of two commuting classical Kac-Moody algebras\footnote{We take a basis of anti-Hermitian generators $T^a$ for the Lie algebra $\mathfrak f$ of $F$ with $[T^a,T^b]=f^{abc}T^c$. We will take the normalisation $\text{Tr}(T^aT^b)=-\delta^{ab}$ in the defining representation. Modes are then defined via $\JJ_\mu^a=\Tr[T^a\JJ_\mu]$ and so are Hermitian $(\JJ^a_\mu)^\dagger=\JJ_\mu^a$.}
\EQ{
\big\{\JJ^a_\pm(x),\JJ^b_\pm(y)\big\}&=f^{abc}\JJ_\pm^c(y)\delta(x-y)\pm\frac {k}{2\pi}\delta^{ab}\delta'(x-y)\ ,\\
\big\{\JJ^a_+(x),\JJ^b_-(y)\big\}&=0\ .
\label{km1}
}
These Poisson brackets are  non ultra-local due to the central term which depends on the derivative of a delta function. As a consequence, they do not give a consistent Poisson bracket for the monodromy matrix: the definition requires a prescription. One way to do this was described by Maillet \cite{Maillet:1985ek}. The problem with the prescription is that it is not clear whether it can be obtained as the classical limit of a quantization of the model.

In the inverse scattering formalism, it is useful to write the Poisson brackets in terms of the spatial component of the Lax connection $\Lax(z)\equiv \Lax_x(z)$. Note that this encodes the whole of the phase space. One way to see this is to notice that
\EQ{
\Lax(z_\pm)=\mp \frac{2\pi}k \JJ_\pm\ ,\qquad z_\pm=\mp\frac{1-\lambda}{1+\lambda}\nu\ .
}
The Poisson bracket can then be written in tensor form as \cite{Hollowood:2015dpa}
\EQ{
&\{\Lax_1(x;z),\Lax_2(y;w)\}=[r_{12}(z,w),\Lax_1(x;z)+\Lax_2(x;w)]\delta(x-y)\\ &\qquad\qquad-[s_{12}(z,w),\Lax_1(x;z)-\Lax_2(y;w)]\delta(x-y)-2s_{12}(z,w)\delta'(x-y)\ .
\label{ikk}
}
The notation is that the bracket acts on a product of $F$ modules $V\otimes V$ and the subscripts indicate which of the copies a quantity acts on. The tensor kernels $r(z,w)$ and $s(z,w)$ act on $V\otimes V$ and are defined as
\EQ{
r(z,w)=\frac{\phi(w)^{-1}+\phi(z)^{-1}}{z-w}\Pi\ ,\qquad s(z,w)=\frac{\phi(w)^{-1}-\phi(z)^{-1}}{z-w}\Pi\ ,
}
where $\Pi$ is the tensor Casimir  operator
\EQ{
\Pi=-\sum_aT^a\otimes T^a
\label{xii}
}
and the ``twist function" is
\EQ{
\phi(z)=\frac{k(1-\lambda^2)(1+\lambda)^2}{2\pi\nu \lambda}\cdot\frac{\nu^2-z^2}{\nu^2(1-\lambda)^2-z^2(1+\lambda)^2}\ .
}

Notice that in this way of formulating the Poisson brackets the non ultra-local term is proportional to the kernel $s(z,w)$ whose non-vanishing relies is implied by the fact that the twist function is non-trivial. A trivial twist function $\phi=$ constant, on the other hand, yields an ultra-local Poisson bracket.

\section{The linear chiral model}\label{s3}

In this section, we define a particular limit of the classical theory that we call the Linear Chiral Model (LCM) and then show that this limiting theory is ultra local. This LCM is the generalization of the $\SU(2)$ case considered by Faddeev and Reshetikhin \cite{Faddeev:1985qu}. The $\SU(2)$ case has also been discussed in the context of the string world sheet in \cite{Klose:2006dd}.

As a classical integrable system parameterized by the pair $(k,\lambda)$, there is an interesting limit in which one takes $k\to0$ and $\lambda\to0$ with the ratio fixed. It is convenient then to fix the free parameter of the Lax connection \eqref{lcc}
\EQ{
\nu=\frac k{4\pi\lambda}\ .
}
Of course this is not a limit that can be reached as a classical limit of the lambda model which requires $k\in Z\to\infty$. 

In this limit, the classical theory has a much simpler structure. The Lax connection becomes
\EQ{
\Lax_\pm(z)=-\frac1{\nu\pm z}\JJ_\pm\ ,
\label{leq}
}
with equations of motion
\EQ{
\partial_\mp\JJ_\pm=\mp\frac1{2\nu  }[\JJ_+,\JJ_-]
\label{jeq}
}
and the Poisson brackets \eqref{km1} loose the central terms:
\EQ{
\big\{\JJ^a_\pm(x),\JJ^b_\pm(y)\big\}&=f^{abc}\JJ_\pm^c(y)\delta(x-y)\ ,\\
\big\{\JJ^a_+(x),\JJ^b_-(y)\big\}&=0\ .
\label{km2}
}
In this limit, the twist function $\phi(z)\rightarrow 2$ and so the Poisson brackets can also be written as a much simpler tensor 
form \eqref{ikk}, 
\EQ{
\{\Lax_1(x;z),\Lax_2(y;w)\}=[r_{12}(z,w),\Lax_1(x;z)+\Lax_2(x;w)]\delta(x-y)\ ,
\label{ikk2}
}
where now
\EQ{
r(z,w)=\frac{\Pi}{z-w}\ .
\label{pip}
}

So the limit leads to a classical theory which is now ultra-local: the $\delta'(x-y)$ term no longer infects the Poisson bracket. This means that the Poisson bracket can be lifted consistently to the monodromy matrix as
\EQ{
\{T_1(z),T_2(w)\}=[r_{12}(z,w),T_1(z)T_2(w)]\ .
\label{mmp}
}

However, the Poisson bracket in the limit becomes degenerate due to the existence of non-trivial centres (quantities that Poisson commute with any other quantity on phase space). These centres are identified with any of the chirally conserved currents 
\EQ{
\Tr[(\JJ_+)^n]\propto \Tr[(A_+)^n]\ ,
} 
including the original Hamiltonian and momentum densities. So in the limit, the original Hamiltonian no longer generates infinitesimal shifts in $t$. In order to recover a consistent phase space, we must impose the constraints
\EQ{
\Tr\big[(\JJ_\pm)^n\big]=\text{const.}
}
Note that these constraints involve $T_{\pm\pm}=$ constant and so are a form of the Pohlmeyer reduction \cite{Miramontes:2008wt}. The limit that we are describing is an example of the alleviation procedure described in  \cite{Delduc:2012qb} which found analogous constraints and the automatic appearance of the Pohlmeyer reduction.

For a general group, the reduced phase space is parameterized as 
\EQ{
\JJ_\pm=g_\pm\Lambda g_\pm^{-1}\ ,
}
for a fixed element of the algebra $\Lambda$. Therefore  the phase space corresponds to a quotient $F/F_0$, where $F_0$ is the stabilizer of $\Lambda$. These space are co-adjoint orbits of the Lie group $F$.\footnote{In the compact case, we can use the inner product $\Tr(T^aT^b)=-\delta^{ab}$  to identify the adjoint and co-adjoint orbits.}
For the $\SU(2)$ case, the quotient is $\SU(2)/\U(1)\simeq S^2$ and we can take
\EQ{
\JJ_+&=-i\BS_+\cdot\Bsigma\ ,\\
\JJ_-&=-i\BS_-\cdot\Bsigma\ ,
}
where
\EQ{
\BS_+\cdot\BS_+=\BS_-\cdot\BS_-=-\frac12\Tr[\Lambda^2]\ .
}

In general, due to a theorem of Borel \cite{Borel}, the co-adjoint orbits are natural homogeneous symplectic---actually K\"ahler---manifolds.\footnote{A nice physicist's review of these spaces appears in \cite{Bordemann:1985xy}.} For generic $\Lambda$ they have the form $F/\U(1)^r$, $r{=}\text{rank}(F)$. But for non-generic $\Lambda$ the group $H_0$ can be larger. The important fact for later is that they can naturally be quantized in terms of certain representations of $F$. For later use, for gauge group $F=\SU(N)$, we will be interested in a particular class of representation,
\EQ{
\Lambda=\Bomega\cdot\BH\ .
\label{gtt}
}
where $\Bomega=k\Be_1$ is the highest weight of the rank $k$ symmetric representation.\footnote{We will see later, that in the quantum theory $k$ will be identified with the level $k$ of the original lambda model.} In this case, the quotient $\SU(N)/U(N-1)\simeq {\mathbb C}P^{N-1}$.\footnote{For $\SU(N)$, we will use an over complete basis of vectors $\Be_i$, with $\Be_i\cdot\Be_j=\delta_{ij}-1/N$ to describe the roots and weights. The roots are $\pm\Be_i\pm\Be_j$, $i\neq j$. The weights of the defining $N$-dimensional representation are $\Be_i$. A representation with highest weight $\Bomega$ will be denoted as $[\Bomega]$. The anti-symmetric representations have highest weights $\Bomega_a=\Be_1+\Be_2+\cdots+\Be_a$. The symmetric representations have highest weights $a\Be_1$. The representations of level $k$ are those for which $(\Be_1-\Be_N)\cdot\Bomega=k$ where $\Be_1-\Be_N$ is the highest root.}

In the limiting theory, the conserved charges $Q_n^{(\pm)}$ are associated to constant currents and so are not dynamically relevant. The Hamiltonian and momentum of the theory now lie in the other set of conserved charges $I_n^{(\pm)}$. Let us extract the first two charges in the series. To achieve this we have to abelianize the connection component $\Lax(z)$. To lowest order $V^{(\pm)}_0=g_\pm$, and so it follows that
\EQ{
\Lax_{-1}^{(\pm)}&=\mp\frac1{2\nu  }\Lambda\ ,\\
\Lax_0^{(\pm)}&=\mathbb P_\BH\Big[g_\pm^{-1} \partial_xg_\pm\mp\frac1{2\nu  }g_+^{-1}g_-\Lambda g_-^{-1}g_+\Big]\ ,
}
where $\mathbb P_\BH$ is a projector on the Cartan subalgebra.

We will identify the light cone components of the energy-momentum as being given by
\EQ{
{\cal P}_\pm&=\pm\Tr\big[\Lambda I^{(\pm)}_0\big]=\pm\int_{-L}^L dx\,\Tr\big[\Lambda \Lax_0^{(\pm)}\big]\\ &=\pm{\cal P}^{(\mp)}+\frac1{2\nu  }\int_{-L}^L dx\,\Tr\big[\JJ_+\JJ_-\big]\ ,
\label{yyr}
}
where
\EQ{
{\cal P}^{(\pm)}=\int_{-L}^Ldx\,\Tr\big[\Lambda g_\pm^{-1} \partial_xg_\pm\big]\ ,
}
generate infinitesimal shifts in $x$:
\EQ{
\{\JJ_\pm,{\cal P}^{(\pm)}\}=\frac12\partial_x\JJ_\pm\ ,\qquad \{\JJ_\mp,{\cal P}^{(\pm)}\}=0\ .
}

For example, for $\SU(2)$, with $\Lambda=is\sigma_3$, $s=k/2$, for each $\BS_\pm=(S_1,S_2,S_3)$, we have
\EQ{
g=(1+|v|^2)^{-1/2}\MAT{1 & \bar v\\ -v&1}\ ,\qquad v=\frac{s-S_3}{S_1+iS_2}\ ,
}
in which case
\EQ{
{\cal P}=\frac1{s}\int_{-L}^L dx\,\frac{S_1\partial_xS_2-S_2\partial_xS_1}{s+S_3}\ ,
}
matching the expression in  \cite{Faddeev:1985qu}.

The Hamiltonian and momentum of the theory can be derived from the following first order action 
\EQ{
S[g_\pm]=\int d^2x\,\Tr\big[\Lambda(g_+^{-1}\partial_-g_++g_-^{-1}\partial_+g_-)+\frac1{2\nu  }g_+\Lambda g_+^{-1}g_-\Lambda g_-^{-1}\big]\ .
}
The LCM is a non-relativistic integrable field theory whose utility lies in the fact that it is ultra local and can be discretized and quantized in a simple way. In short, we can apply the QISM to it.

\section{Quantum inverse scattering method}\label{s4}

In order to apply the QISM to the LCM model, we first define a discrete version of the theory in space. In fact, it turns out that it is more natural define a discrete version of the whole of 2d Minkowski spacetime rather than just space. The context here is  the light-cone approach to integrable field theories \cite{Faddeev:1985qu,Destri:1987ze,Destri:1987hc,Destri:1987ug,Faddeev:1996iy}.

\begin{figure}
\begin{center}
\begin{tikzpicture}[scale=0.7]
\node at (-2,0) {\footnotesize $\JJ_{-,n-1}$};
\node at (0,0) {\footnotesize $\JJ_{+,n-1}$};
\node at (2,0) {\footnotesize $\JJ_{-,n}$};
\node at (4,0) {\footnotesize $\JJ_{+,n}$};
\node at (6,0) {\footnotesize $\JJ_{-,n+1}$};
\node at (8,0) {\footnotesize $\JJ_{+,n+1}$};
\draw[gray] (-2,-2) -- (2,2);
\draw[gray] (2,-2) -- (6,2);
\draw[gray] (6,-2) -- (9,1);
\draw[gray] (0,-2) -- (-3,1);
\draw[gray] (4,-2) -- (0,2);
\draw[gray] (8,-2) -- (4,2);
\node at (-1,-3) {\footnotesize $(n-1)\Delta$};
\node at (3,-3) {\footnotesize $n\Delta$};
\node at (7,-3) {\footnotesize $(n+1)\Delta$};
\begin{scope}[xshift=-4cm,yshift=-1.5cm]
\draw[very thick,<->] (-0.5,-0.5) -- (-0.5,-1.5) -- (0.5,-1.5);
\node at (0.8,-1.5) {$x$};
\node at (-0.5,-0.2) {$t$};
\end{scope}
\begin{scope}[xshift=9.6cm,yshift=-1.5cm]
\draw[very thick,<->,rotate=45] (-0.5,-0.5) -- (-0.5,-1.5) -- (0.5,-1.5);
\node at (1.8,-0.2) {$x^+$};
\node at (0,-0.2) {$x^-$};
\end{scope}
\end{tikzpicture}
\caption{\footnotesize  The null lattice in spacetime. The degrees of freedom live along the null segments as indicated.}
\label{f2} 
\end{center}
\end{figure}
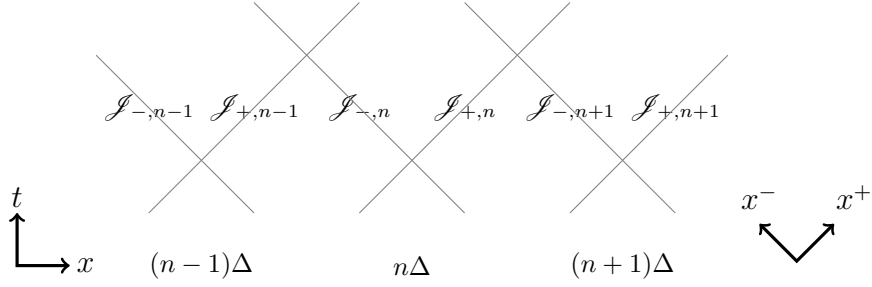

The discrete theory in spacetime is defined on a light-cone lattice associated to the points
\EQ{
x^+=n\Delta\ ,\qquad x^-=m\Delta\ ,\qquad m,n\in\mathbb Z\ ,
}
illustrated in fig.~\ref{f2}.
The degrees-of-freedom, or spins, in the form of the discrete modes  $\JJ_{\pm,n}^a$ lie on the null links of the lattice as shown in the figure. So on an equal time slice, say $t=0$, each lattice point $x=n\Delta$ is associated to a pair of modes $\JJ^a_{\pm,n}$---actually located on the null links $x^-=-n\Delta$ and $x^+=n\Delta$, respectively---and we impose periodic boundary conditions $\JJ^a_{\pm,n+p}\equiv\JJ^a_{\pm,n}$. 

In lattice models gauge fields live on links of the lattice and so it is perfectly natural that the spins here are located on the links because in the classical theory they define the Lax connection.

\ST{Quantization}

We can quantize the discrete LCM model by replacing the Poisson brackets \eqref{km2} with commutators $\{\ ,\ \}\rightarrow -i[\ ,\ ]$
\EQ{
[\JJ_{\pm,m}^a\JJ_{\pm,n}^b]=\frac i\Delta f^{abc}\JJ^c_{\pm,n}\delta_{mn}\ .
}
The operators can be represented by generators in a particular representation ${\cal R}$ of $\mathfrak f$,
\EQ{
\JJ_{+,n}^a=-\frac i\Delta T_{2n}^a\ ,\qquad \JJ_{-,n}^a=-\frac i\Delta T^a_{2n-1}\ .
} 
So the Hilbert space is a product $V_1\otimes V_2\otimes\cdots \otimes V_{2p}$, where $V$ is the module for the representation ${\cal R}$. Therefore in the quantum theory $\Lambda$ in \eqref{gtt} is quantized in the sense that $\Bomega$ must be a weight vector that we can choose to be a highest weight vector. Each choice of representation gives a different quantization of the classical model. In our application to the lambda model we will need a specific choice of ${\cal R}$, namely the rank $k$ symmetric representation.

The currents take the form
\EQ{
\JJ_{+,n}&=\sum_a\JJ_{+,n}^aT^a=-\frac i\Delta \sum_aT^a_{2n}\otimes T^a=\frac i\Delta\Pi_{2n,0}\ ,\\
\JJ_{-,n}&=\sum_a\JJ_{-,n}^aT^a=-\frac i\Delta \sum_aT^a_{2n-1}\otimes T^a=\frac i\Delta\Pi_{2n-1,0}\ ,
}
acting on $V_{2n}\otimes V_0$ and $V_{2n-1}\otimes V_0$, respectively, where $V_{n}$ is the $n^\text{th}$ factor in the Hilbert space and $V_0$ can be viewed as an auxiliary space representing the underlying original Lie algebra structure of the Lax equations \eqref{leq} and \eqref{jeq}.

The spins are associated to the infinitesimal monodromy over the null links
\EQ{
&\text{P}\overset{\longleftarrow}{\text{exp}}\Big[-\int_{n\Delta}^{(n+1)\Delta}dx^+\,\Lax_+(x;z)\Big]\thicksim R_{2n,0}(z+\nu)\ ,\\
&\text{P}\overset{\longleftarrow}{\text{exp}}\Big[-\int_{n\Delta}^{(n-1)\Delta}dx^-\,\Lax_-(x;z)\Big]\thicksim R_{2n-1,0}(z-\nu)\ ,
\label{bxx}
}
where $R(z)$ is the ubiquitous ``$R$ matrix" associated the representation of ${\cal R}$ of $F$ acting on $V\otimes V$. Note that in the following we will choose the auxiliary space $V_0$ to be the module for the same representation ${\cal R}$. 

Integrability of the lattice model is ensured if the $R$-matrix satisfies the Yang-Baxter equation
\EQ{
R_{12}(z-w)R_{n,1}(z)R_{n,2}(w)=R_{n,2}(w)R_{n,1}(z)R_{12}(z-w)\ .
\label{ybe}
}
Note that in the tensor notation we have two copies of the auxiliary space labelled $1$ and $2$ here. 
The $R$ matrix also satisfies the fundamental regularity property
\EQ{
R(0)=P\ ,
\label{reg}
}
where $P$ permutes the two modules on which $R(0)$ acts. 

For the $R$ matrix, the classical limit involves taking $z\to\infty$, \cite{Destri:1987hc}
\EQ{
R(z)\longrightarrow 1+\frac{i\Pi+\gamma}z+\cdots\ .
\label{scl}
}
Here, $\Pi$ is tensor Casimir \eqref{xii} and $\gamma$ is an unimportant constant that depends on the representation. This ensures that the relations \eqref{bxx}  give the spins in the semi-classical limit:
\EQ{
&\text{P}\overset{\longleftarrow}{\text{exp}}\Big[-\int_{n\Delta}^{(n+1)\Delta}dx^+\,\Lax_+(x;z)\Big]\thicksim R_{2n,0}(z+\nu)=1+\frac{\Delta\JJ_{+,n}+\gamma}z+\cdots\ ,\\
&\text{P}\overset{\longleftarrow}{\text{exp}}\Big[-\int_{n\Delta}^{(n-1)\Delta}dx^-\,\Lax_-(x;z)\Big]\thicksim R_{2n-1,0}(z-\nu)=1+\frac{\Delta\JJ_{-,n}+\gamma}z+\cdots\ ,
\label{byy2}
}

\ST{Commutation relations and Poisson brackets}

In order to recover the Poisson brackets in the classical limit, it is necessary to define the
the monodromy over one spatial step of the lattice,
\EQ{
T_{n,0}(z)=R_{2n,0}(z+\nu)R_{2n-1,0}(z-\nu)\ ,
}
It then follows that in the classical limit (ignoring the constants)
\EQ{
T_{n,0}(z)\longrightarrow 1+\frac{i\Pi_{2n-1,0}}{z+\nu}+\frac{i\Pi_{2n,0}}{z-\nu}+\cdots
}
which we identify with 
\EQ{
\text{P}\overset{\longleftarrow}{\text{exp}}\Big[-\int_{n\Delta}^{(n+1)\Delta}dx\,\Lax(x;z)\Big]=1-\Delta\Lax_n(z)+\cdots\ .
}

The fundamental commutation relations of the theory are defined in terms of the single step monodromy matrices $T_n(z)$ which follow from the Yang-Baxter equation \eqref{ybe}
\EQ{
R_{12}(z-w)T_{n,1}(z)T_{n,2}(w)=T_{n,2}(w)T_{n,1}(z)R_{12}(z-w)\ .
\label{fcr}
}
In order to take the classical limit, we write this as a commutation relation,
\EQ{
[T_{n,1}(z),T_{n,2}(w)]&=\big(1-R_{12}(z-w)\big)T_{n,1}(z)T_{n,2}(w)\\ &\qquad\qquad-T_{n,2}(w)T_{n,1}(z)\big(1-R_{12}(z-w)\big)\ .
}
Taking the classical limit, we have
\EQ{
[T_{n,1}(z),T_{n,2}(w)]\longrightarrow -i\Delta^2\{\Lax_{n,1}(z),\Lax_{n,2}(w)\}\ ,
}
on the left-hand side, and given \eqref{scl} we have
\EQ{
R(z)\longrightarrow1+ir(z)+\frac\gamma z+\cdots\ ,
}
the right-hand side becomes
\EQ{
-i\Delta[r_{12}(z-w),\Lax_{n,1}(z)+\Lax_{n,2}(w)]\ .
}
Hence, in the limit we have
\EQ{
\{\Lax_{n,1}(z),\Lax_{n,2}(w)\}= \frac1\Delta[r_{12}(z-w),\Lax_{n,1}(z)+\Lax_{n,2}(w)]\ ,
}
which is a discrete version of the Poisson bracket algebra of the LCM, so  \eqref{ikk2} with \eqref{pip}.

\begin{figure}
\begin{center}
\begin{tikzpicture}[scale=0.7]
\node at (-2,0) {\footnotesize $\JJ_{-,n-1}$};
\node at (0,0) {\footnotesize $\JJ_{+,n-1}$};
\node at (2,0) {\footnotesize $\JJ_{-,n}$};
\node at (4,0) {\footnotesize $\JJ_{+,n}$};
\node at (6,0) {\footnotesize $\JJ_{-,n+1}$};
\node at (8,0) {\footnotesize $\JJ_{+,n+1}$};
\draw[gray] (-2,-2) -- (2,2);
\draw[gray] (2,-2) -- (6,2);
\draw[gray] (6,-2) -- (9,1);
\draw[gray] (0,-2) -- (-3,1);
\draw[gray] (4,-2) -- (0,2);
\draw[gray] (8,-2) -- (4,2);
\draw[very thick] (-2.25,0.25) --  (-3,1);
\draw[very thick] (-1.75,-0.25) -- (-1,-1) -- (-0.25,-0.25);
\draw[very thick] (0.25,0.25) -- (1,1) -- (1.75,0.25);
\draw[very thick] (2.25,-0.25) -- (3,-1) -- (3.75,-0.25);
\draw[very thick] (4.25,0.25) -- (5,1) -- (5.75,0.25);
\draw[very thick] (6.25,-0.25) -- (7,-1) -- (7.75,-0.25);
\draw[very thick] (8.25,0.25) -- (9,1);
\begin{scope}[xshift=-4cm,yshift=-1.5cm]
\draw[very thick,<->] (-0.5,-0.5) -- (-0.5,-1.5) -- (0.5,-1.5);
\node at (0.8,-1.5) {$x$};
\node at (-0.5,-0.2) {$t$};
\end{scope}
\begin{scope}[xshift=9.6cm,yshift=-1.5cm]
\draw[very thick,<->,rotate=45] (-0.5,-0.5) -- (-0.5,-1.5) -- (0.5,-1.5);
\node at (1.8,-0.2) {$x^+$};
\node at (0,-0.2) {$x^-$};
\end{scope}
\end{tikzpicture}
\caption{\footnotesize  The sawtooth contour in space time that defines the monodromy.}
\label{f3} 
\end{center}
\end{figure}
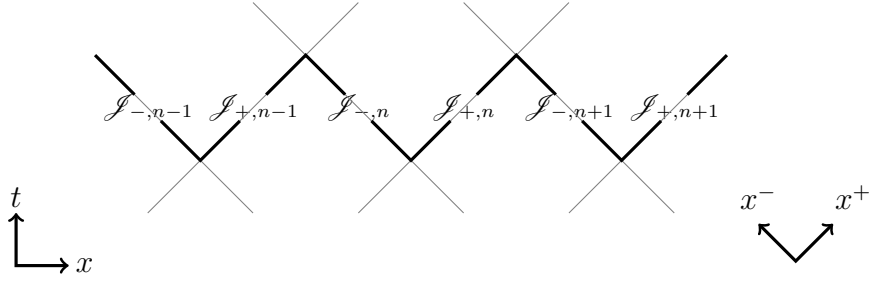

In the QISM the total monodromy matrix (acting on the auxiliary space $V_0$) whose elements are operators on the Hilbert space is given by
\EQ{
T(z)=T_p(z)T_{p-1}(z)\cdots T_1(z)=R_{2p,0}(z+\nu)R_{2p-1,0}(z-\nu)\cdots R_{1,0}(z-\nu)\ .
\label{juw}
}
This is the monodromy matrix of an inhomgeneous spin chain with alternating inhomogeneities $(-1)^n\nu$. We can identify it with a discrete version of the continuum monodromy integrated along the sawtooth contour in spacetime shown in  fig.~\ref{f3}.

The fundamental commutation relations \eqref{fcr} ensure that 
\EQ{
[\Tr_0T(z),\Tr_0T(w)]=0
\label{iuu}
}
and so $\Tr_0T(z)$ provides a generating function for the conserved quantities of the discrete theory including the energy and momentum.

\ST{Energy and momentum}

In the light cone lattice approach, we will identify the light cone components of the energy and momentum as \cite{Faddeev:1985qu,Destri:1987hc}
\EQ{
U_+=e^{-i\Delta{\cal P}_+}=\Tr_0T(\nu)\ ,\qquad U_-^\dagger=e^{i\Delta{\cal P}_-}=\Tr_0T(-\nu)\ .
}
These unitary operators generate shifts on the light cone lattice $x^+\to x^++\Delta$ and $x^-\to x^--\Delta$. Note that $U_+$ commutes with $U_-$ on account of \eqref{iuu}.

Let us consider the expression for $U_+$ in more detail. Using the regularity condition \eqref{reg}, we have 
\EQ{
U_+&=\Tr_0\big[R_{2p,0}(2\nu)P_{2p-1,0}R_{2p-2,0}(2\nu)\cdots P_{1,0}\big]\\
&=\Omega_+R_{2p,2p-1}(2\nu  )R_{2p-2,2p-3}(2\nu  )\cdots R_{2,1}(2\nu  )\ ,
}
where 
\EQ{
\Omega_+=\Tr_0\big[P_{2p-1,0}P_{2p-3,0}\cdots P_{1,0}\big]=P_{1,3}P_{3,5}\cdots P_{2p-3,2p-1}\ ,
}
is an operator which cyclically permutes the odd spaces (spatial shift in the $+x$ direction):
\EQ{
\Omega_+^{-1}T^a_{2n-1}\Omega_+=T^a_{2n+1}\ .
}

Now we take the take the classical limit, $U_+\to1-i\Delta{\cal P}_++\cdots$ and use \eqref{scl} to discover that in this limit
\EQ{
{\cal P}_+={\cal P}^{(-)}+\frac\Delta{2\nu  }\sum_{n=1}^p\Tr\big[\JJ_{+,n}\JJ_{-,n}\big]+\cdots
}
which we identify with a discretization of \eqref{yyr}. We have identified the shift operator with $\Omega_+=\exp(-i\Delta {\cal P}^{(-)})$. There is also a constant contribution that plays no r\^ole so we have ignored it.

There is a similar story for $U_-$:
\EQ{
U_-^\dagger&=\Tr_0\big[P_{2p,0}R_{2p-1,0}(-2\nu)P_{2p-2,0}\cdots R_{1,0}(-2\nu)\big]\\
&=R_{2p-1,2p}(-2\nu)R_{2p-3,2p-2}(-2\nu )\cdots R_{1,2}(-2\nu)\Omega_-\ ,
}
where
\EQ{
\Omega_-=\Tr_0\big[P_{2p,0}P_{2p-2,0}\cdots P_{2,0}\big]=P_{2,4}P_{4,6}\cdots P_{2p-2,2p}\ ,
}
is an operator which cyclically permutes the even spaces:
\EQ{
\Omega_-^{-1}T^a_{2n+2}\Omega_-=T^a_{2n}\ .
}

Following the steps as above in the classical limit we get
\EQ{
{\cal P}_-=-{\cal P}^{(+)}+\frac\Delta{2\nu  }\sum_{n=1}^p\Tr\big[\JJ_{+,n}\JJ_{-,n}\big]+\cdots
}
which is a discretization of \eqref{yyr}.

We can therefore express the energy (Hamiltonian) and momentum in terms of the trace of the monodromy matrix
\EQ{
{\cal E}\equiv H=-\frac i\Delta\log\frac{\Tr_0T(\nu)}{\Tr_0 T(-\nu)}\ ,\qquad {\cal P}=\frac i\Delta\log\big[\Tr_0 T(\nu)\Tr_0 T(-\nu)\big]\ .
}
Note that since $R_{ab}(\nu)R_{ba}(-\nu)=1$, the momentum operator generates---as it must---a spatial shift in the lattice, for the odd and even modes separately, 
\EQ{
e^{-i\Delta{\cal P}}=\Omega_+\Omega_-=P_{1,3}P_{3,5}\cdots P_{2p-3,2p-1}P_{2,4}P_{4,6}\cdots P_{2p-2,2p}\ .
}

\begin{figure}
\begin{center}
\begin{tikzpicture}[scale=0.7,decoration={markings,mark=at position 0.55 with {\arrow{>}}}]
\draw[very thick,postaction={decorate}] (2,-2) -- (4,0);
\draw[very thick,postaction={decorate}] (0,0) -- (2,-2);
\draw[very thick,postaction={decorate}] (2,2) -- (4,0);
\draw[very thick,postaction={decorate}] (0,0) -- (2,2);
\node at (5,-1.5) {$R_{2n,0}(z+\nu)$};
\node at (-1,-1.5) {$R_{2n-1,0}(z-\nu)$};
\node at (6,1.5) {$U_+^{-1}R_{2n-1,0}(z-\nu)U_+$};
\node at (-2,1.5) {$U_-^{-1}R_{2n,0}(z+\nu)U_-$};
\end{tikzpicture}
\caption{\footnotesize  The discrete version of the flatness condition for an elementary plaquette.}
\label{f4} 
\end{center}
\end{figure}
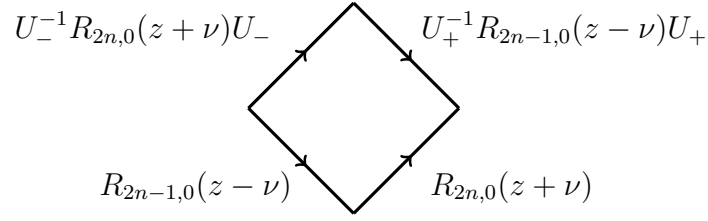

\ST{Classical equations of motion}

It is interesting to show that the classical equations motion can be derived directly from the quantum lattice model in the appropriate limit \cite{Faddeev:1996iy,Destri:1987hc}. The classical equations of motion take the form of a flatness condition that ensures that the monodromy is path independent. A lattice version of this can be formulated for a single plaquette on the lattice, shown in fig.~\ref{f4}. The analogue of flatness can be expressed by the identity
\EQ{
R_{2n,0}(z+\nu)R_{2n-1,0}(z-\nu)=U_+^{-1}R_{2n-1,0}(z-\nu)U_+U_-^{-1}R_{2n,0}(z+\nu)U_-\ .
}
It is straightforward to show that in the classical limit \eqref{byy2}, this becomes
\EQ{
\frac1{z+\nu}\frac1\Delta\big(U_-^{-1}\JJ_{+,n}U_--\JJ_{+,n}\big)+&
\frac1{z-\nu}\frac1\Delta\big(U_+^{-1}\JJ_{-,n}U_+-\JJ_{-,n}\big)\\ &-\frac1{z^2-\nu^2}[\JJ_{+,n},\JJ_{-,n}]=0\ ,
}
which gives rise to
\EQ{
\frac1{z+\nu}\partial_-\JJ_{+}+
\frac1{z-\nu}\partial_+\JJ_{-}-\frac1{z^2-\nu^2}[\JJ_+,\JJ_-]=0\ ,
}
in the continuum limit. This is just the classical Lax equation \eqref{leq}.

\ST{Symmetric representations}

For our application to the lambda model, we will focus on the group $F=\SU(N)$. The relevant representation ${\cal R}$ for the spin chain will turn out to be the rank-$k$ symmetric representation. For the fundamental representation $k=1$, we have
\EQ{
R(z)=\frac{iz+P}{iz+1}\ ,
}
where
\EQ{
P=-\Pi+\frac1N\ ,
}
is the permutation operator.

The $R$ matrix for the higher rank symmetric representations with highest weight $k\Be_1$ can be obtained by the fusion procedure \cite{Kulish:1981gi}. One way to write the result is in terms of the projectors onto the representations that appear in the tensor product of two symmetric representations:
\EQ{
[k\Be_1]\times[k\Be_1]=\sum_{j=0}^k[(2k-j)\Be_1+j\Be_2]\ .
}
in the form
\EQ{
R(z)=\sum_{j=0}^k\rho_j(z)\mathbb P_{(2k-j)\Be_1+j\Be_2}\ ,
}
with
\EQ{
\rho_j(z)=\prod_{\ell=0}^{j-1}\frac{z-(k-\ell)}{z+(k+\ell)}\ .
}

The example $F=\SU(2)$ with spins in the spin $s=k/2$ representation is precisely the spin model constructed by Faddeev and Reshetikhin \cite{Faddeev:1985qu}.

\section{Algebraic Bethe Ansatz}\label{s5}

The beauty of formulating the discrete model in the way we have done is that the eigenvectors and eigenvalues of the null shift operators $\U_\pm$ and hence the energy and momentum can be found exactly by the Algebraic Bethe Ansatz (ABA). A complete review of these techniques would be unnecessary and lengthy (we refer to the review \cite{Faddeev:1996iy} and book \cite{KorepinBook}), so here we limit ourselves to the bare bones and only give explicit expressions for $F=\SU(2)$.

The ABA provides a formalism to construct the simultaneous eigenvectors of the general class of transfer matrices
\EQ{
T(z)=R_{2p,0}(z-\nu_{2p})R_{2p-1,0}(z-\nu_{2p-1})\cdots R_{1,0}(z-\nu_1)\ ,
}
for arbitrary {\it inhomogeneities\/} $\{\nu_n\}$ and for arbitrary $z$. In our case, inhomogeneities are alternating $\nu_n=(-1)^n\nu$. It is also possible to have different spins on each lattice site, but here the spins are all associated to the same representation ${\cal R}$.

For $\SU(2)$, let us write
\EQ{
T(z)=\MAT{A(z) & B(z)\\ C(z) & D(z)}\ ,
}
where $A(z)$, etc, are operators on the Hilbert space. The eigenstates are given by the vectors
\EQ{
\ket{\Psi(z_1,\ldots,z_m)}=B(z_1)B(z_2)\cdots B(z_m)\ket{\Omega}\ ,
\label{ter}
}
where $\ket{\Omega}$ is a reference state, the ``pseudo vacuum", taking the form $\ket{\Omega}=\ket{\uparrow\uparrow\cdots\uparrow}$. This is the ground state of the ferromagnetic spin chain. Here, we are in the anti-ferromagnetic regime and so $\ket{\Omega}$ is not the true ground state. 

The states \eqref{ter} are eigenstates if the parameters $\{z_i\}$ satisfy the celebrated Bethe Ansatz Equations (BAE)
\EQ{
\Big(\frac{z_i+\nu +ik/2}{z_i+\nu -ik/2}\Big)^p\Big(\frac{z_i-\nu +ik/2}{z_i-\nu -ik/2}\Big)^p=\prod_{j=1\atop(\neq i)}^m\frac{z_i-z_j+i}{z_i-z_j-i}\ .
}
Note that when the inhomogeneities vanish $\nu=0$, these are precisely the BAE of the Heisenberg $XXX_{k/2}$ spin chain.  We will find that the presence of the inhomogeneities $\pm\nu $ affects some quantities but the overall structure of the solutions is unaffected.

The eigenvalues of the null evolution  operators are
\EQ{
U_\pm=\prod_{j=1}^m\frac{z_j\mp\nu \pm ik/2}{z_j\mp\nu \mp ik/2}\ .
}
The energy and momentum are then given as sums over contributions from each Bethe root $z_j$:
\EQ{
{\cal E}=\sum_{j=1}^m\varepsilon_0(z_j)\ ,\qquad {\cal P}=\sum_{j=1}^m\wp_0(z_j)\ .
}
Each root $z_j$ is related to a pseudo particle excitation with energy and momentum
\EQ{
\varepsilon_0(z)&=\frac2\Delta\Big(\tan^{-1}\Big[\frac2k(z-\nu)\Big]-\tan^{-1}\Big[\frac2k(z+\nu)\Big]-\pi\Big)\ ,\\
\wp_0(z)&=\frac2\Delta\Big(\tan^{-1}\Big[\frac2k(z-\nu)\Big]+\tan^{-1}\Big[\frac2k(z+\nu)\Big]\Big)\ ,
\label{ppe}
}
with the branches of the functions chosen appropriately. The parameter $z$ is a kind of rapidity variable.

For other groups, the construction generalizes: the eigenstates are given in terms of vectors which depend on parameters that satisfy auxiliary equations, the now more complicated nested BAE. 

\subsection{The ground state}

The energy of a pseudo particle is negative and so the true ground state will involve filling the pseudo vacuum with pseudo particles. In the thermodynamic limit, the Bethe roots are known to group into strings, for which
\EQ{
z_{\alpha}=z+\frac i2(M+1-2\alpha)\ ,\qquad \alpha=1,2,\ldots,M\ .
\label{snn}
}
The true ground state is then associated with a configuration of Bethe roots in the form of a condensate of $k$ strings matching the rank of representation ${\cal R}$. 

In the thermodynamic limit, the density of $k$ strings in the vacuum is determined by an integral equation that results from taking the continuum limit of the BAE:
\EQ{
\rho^{(k)}(z)+\frac1\pi\int_{-\infty}^\infty K(z-w)\rho^{(k)}(w)dw=-\frac1{2\pi}\frac{d\wp_0^{(k)}(z)}{dz}\ ,
}
where $\wp^{(k)}_0(z)$ is the momentum of a $k$ string obtained by summing over the momenta of pseudo particles in the set \eqref{snn} with $M=k$. The kernel in the above is obtained by averaging the derivative of the basic scattering phase over the strings:
\EQ{
K(z-w)&=i\frac d{dz}\sum_{\alpha,\beta=1}^k\log\frac{z_\alpha-w_\beta+i}{z_\alpha-w_\beta-i}\\ &=\sum_{\alpha=1}^{k-1}\frac{4\alpha}{(z-w)^2+\alpha^2}+\frac{2k}{(z-w)^2+k^2}\ .
\label{dky}
}
Solving the integral equation (via Fourier Transform) gives the density of $k$ strings in the ground state
\EQ{
\rho^{(k)}(z)=\frac1{4\cosh \pi(z-\nu)}+\frac1{4\cosh\pi(z+\nu)}\ ,
\label{vst}
}
an expression that depends directly on the coupling $\nu$. Notice, however, that the density does not depend on the rank $k$.

The density of strings in the ground state in the limit of vanishing inhomogeneity $\nu=0$ and spin $\frac12$ is precisely the Hulth\'en solution\footnote{An excellent summary of the Heisenberg spin chain is the book by Takahashi \cite{Tak}.} of the anti-ferromagnetic $XXX$ Heisenberg spin chain. It is noteworthy that the ground state is non trivial in the sense of having  non-trivial entanglement characteristic of the vacuum state of a relativistic QFT.

\subsection{The excitations}

From the point of view of the QFT, we are interested in the spectrum of single particle states above the ground state. For the spin chain around its anti-ferromagnetic ground state this was solved some time ago (again we refer to the book \cite{Tak}). However, the interpretation in terms of particles is somewhat subtle \cite{TF,FT}: the excitations have a hidden kink structure which means that there are non-trivial selection rules given that we are working with periodic boundary conditions. 

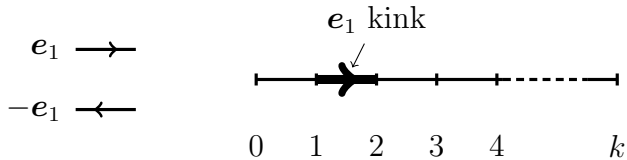
\begin{figure}[ht]
\begin{center}
\begin{tikzpicture} [scale=0.8,decoration={markings,mark=at position 0.7 with {\arrow{>}}}]
\draw[very thick] (0,0) -- (4.1,0);
\draw[very thick,densely dashed] (4.1,0) -- (5.5,0);
\draw[very thick] (5.5,0) -- (6,0);
\draw[very thick] (0,-0.1) -- (0,0.1);
\draw[very thick] (1,-0.1) -- (1,0.1);
\draw[very thick] (2,-0.1) -- (2,0.1);
\draw[very thick] (3,-0.1) -- (3,0.1);
\draw[very thick] (4,-0.1) -- (4,0.1);
\draw[very thick] (6,-0.1) -- (6,0.1);
\draw[line width=1.2mm,postaction={decorate}] (1,0) -- (2,0);
\node at (0,-1.1) {$0$};
\node at (1,-1.1) {$1$};
\node at (2,-1.1) {$2$};
\node at (3,-1.1) {$3$};
\node at (4,-1.1) {$4$};
\node at (6,-1.1) {$k$};
\begin{scope}[xshift=-3cm,yshift=0.5cm];
\draw[postaction={decorate},very thick] (0,0) -- (1,0);
\draw[postaction={decorate},very thick] (1,-1) -- (0,-1);
\node at (-0.5,0) {$\Be_1$};
\node at (-0.7,-1) {$-\Be_1$};
\end{scope}
\node at (2,1) (a1) {$\Be_1$ kink};
\draw[->] (a1) -- (1.6,0.2);
\end{tikzpicture}
\end{center}
\caption{The hidden RSOS kink structure of the spinons. The kinks have a charge associated to weights of the spin $\frac12$ representations $\pm\Be_1$ and the vacua are the set of highest weights at level $\leq k$.}
\label{f5}
\end{figure}

For the $\SU(2)$ case, with the spin $k/2$ representation, the kink structure requires that the excitations---the spinons (or Cloizeaux-Pearson modes)---appear in pairs. They correspond to holes in the distribution of $k$ strings. It is a non-trivial problem to determine their structure because of the ``back flow" on the $k$ strings themselves when a hole is made in the distribution. Each spinon transforms as a doublet under $\SU(2)$ but has, in addition, a hidden kink nature that can be described as an RSOS (restricted solid on solid) structure at level $k$ \cite{Reshetikhin:1990jn,Faddeev:1996iy}. This means that the kinks are associated to a set of $k+1$ vacua labelled by the $\SU(2)$ highest weights at level $\leq k$, so the vectors $j\Be_1$, $j=0,1,2\ldots,k$. The kinks $K_{ab}$ then correspond to a pair of vacua connected by  a basic link on the diagram illustrated in fig.~\ref{f5}.  

The kinks have topological charge that is $\pm\Be_1$ corresponding to the two representations that generically appear in the tensor product
\EQ{
[\Be_1]\times[\Bomega]=[\Bomega+\Be_1]+[\Bomega-\Be_1]\ ,
}
or in terms of spins $[\frac12]\times[j]=[j+\frac12]+[j-\frac12]$. Note that in periodic boundary conditions, states can only contain an even number of spinons half with kink charge $+\Be_1$ and half with $-\Be_1$. More general states should be obtained by introducing non-trivial boundary conditions for the chain.

Even though there are selection rules on spinon states resulting from the kink structure, we can still view them as individual excitations with their own energy and momentum. Effectively, one can assign each spinon a dispersion relation
\EQ{
\varepsilon(z)&=\frac2\Delta\tan^{-1}\Big(\frac{\cosh\pi z}{\sinh \pi\nu }\Big)\ ,\qquad
\wp(z)=\frac2\Delta\tan^{-1}\Big(\frac{\sinh\pi z}{\cosh\pi\nu }\Big)\ ,\\
\text{i.e.}\qquad &\sinh^2\pi\nu\tan^2\frac{\Delta\varepsilon}2-\cosh^2\pi\nu\tan^2\frac{\Delta\wp}2=1\ .
\label{xen}
}
These excitations have a gap $2\Delta^{-1}\tan^{-1}(1/\sinh \pi\nu )$. Note that the excitation has $\Delta|\wp|\leq\pi$.

It is interesting to compare the above with the energy and momentum of the same excitations in the $XXX_{k/2}$ Heisenberg spin chain. In this model, the momentum is same as the above with $\nu=0$ and the energy is given by $2^{-1}dp(z)/dz$:
\EQ{
\text{(Heisenberg)}\qquad\varepsilon(z)&=\frac\pi\Delta\sech\pi z\ ,\qquad \wp(z)= \frac2\Delta\tan^{-1}(\sinh\pi z)\ ,\\
\text{i.e.}\qquad \varepsilon&=\frac\pi\Delta\cos\frac{\Delta\wp}2\ ,
\label{xxe}
}
with $\Delta|\wp|\leq\pi$.
So in contrast with the light cone lattice model, the spinons in the $XXX_{k/2}$ Heisenberg spin chain are gapless. We will discuss the spin chain point of view in section \ref{s5.3}.

Returning to \eqref{xen}, what is particularly interesting is that there exists a non-trivial continuum limit where $\Delta\to0$ and $\nu\to\infty$ with 
\EQ{
\frac1\Delta e^{-\pi\nu }=\frac m4
\label{clm}
} 
fixed. So in this limit, a mass scale $m$ emerges and we obtain a relativistic dispersion relation with $\theta=\pi z$ being the relativistic rapidity:
\EQ{
&\varepsilon(\theta)=m\cosh \theta\ ,\qquad \wp(\theta)=m\sinh\theta\ ,\\
&\text{i.e.}\qquad \varepsilon^2-\wp^2=m^2\ .
}
This provides a concrete example of the phenomenon of {\it dimensional transmutation\/} in QFT, where a mass scale is generated out of a cut off, here $\mu=\Delta^{-1}$, and a dimensionless coupling, here $\nu$. In particular, the beta function of the coupling---giving the way $\nu$ must vary with the cut off to keep the mass scale $m$ fixed---is
\EQ{
\mu\frac{d\nu}{d\mu}=\frac 1\pi\ .
}
Given that $\nu=k/(4\pi\lambda)$, this is precisely the beta function of the $\SU(2)$ lambda model \eqref{bft} in the UV limit $\lambda\to0$ (at large $k$).

It is remarkable that the discussion generalizes to an arbitrary group although the details are a good deal more complicated. 
The generalization of the Heisenberg chain to $\SU(N)$ with spins in the symmetric representation was discussed by Johanesson \cite{Johannesson:1986ig} and more generally by Destri and de Vega \cite{Destri:1987hc}.

The ground state again consistent of $k$ strings, although now the strings carry a branch label to reflect the higher rank group structure, $a=1,2,\ldots,r=N-1$, for $\SU(N)$. Once again the excitations correspond to hole in the distribution of $k$ strings in each branch. In order to determine the spectrum, we only need the eigenvalues of the transfer matrix in the thermodynamic limit and for large $\nu$ as given in  \cite{Destri:1987hc}. One finds that the excitation in the $a^\text{th}$ branch has 
\EQ{
\varepsilon(\theta)=\frac{\lambda_a}\Delta e^{-2\pi\nu /c_2(F)}\cosh\theta\ ,\qquad\wp(\theta)=\frac{\lambda_a}\Delta e^{-2\pi\nu /c_2(F)}\sinh\theta\ ,
}
where $\theta=2\pi z/c_2(F)$ and $\lambda_a$ are a characteristic set of numbers for each group. For $\SU(N)$,
\EQ{
\lambda_a=\sin\frac{\pi a }N\ ,\qquad a=1,2,\ldots,N-1\ .
}
Taking a continuum limit, $\Delta\to0$ and $\nu\to\infty$ with 
\EQ{
\frac1\Delta\exp\Big[-\frac{2\pi\nu}{c_2(F)}\Big]=m\ ,
\label{msl}
}
fixed, we find a relativistic spectrum of exictations with masses
\EQ{
m_a=m\sin\frac{\pi a}N\ ,
}
for $\SU(N)$. These states transform in the $a^\text{th}$ anti-symmetric representation of $\SU(N)$, i.e.~with highest weight $\Be_1+\Be_2+\cdots+\Be_a$.

It is important to emphasize that the spin chain is built from a {\it symmetric\/} representation while the excitations correspond to the {\it anti-symmetric\/} representations.

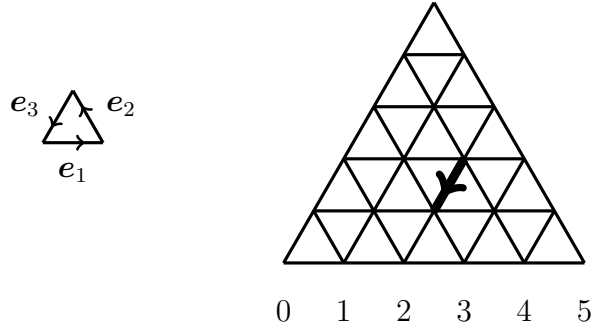
\begin{figure}[ht]
\begin{center}
\begin{tikzpicture} [scale=0.8,decoration={markings,mark=at position 0.7 with {\arrow{>}}}]
\draw[very thick] (0,0) -- (5,0);
\draw[very thick] (0.5,0.866) -- (4.5,0.866);
\draw[very thick] (1,1.732) -- (4,1.732);
\draw[very thick] (1.5,2.598) -- (3.5,2.598);
\draw[very thick] (2,3.464) -- (3,3.464);
\draw[very thick] (0,0) -- (2.5,4.33);
\draw[very thick] (5,0) -- (2.5,4.33);
\draw[very thick] (1,0) -- (0.5,0.866);
\draw[very thick] (2,0) -- (1,1.732);
\draw[very thick] (3,0) -- (1.5,2.598);
\draw[very thick] (4,0) -- (2,3.464);
\draw[very thick] (1,0) -- (3,3.464);
\draw[very thick] (2,0) -- (3.5,2.598);
\draw[very thick] (3,0) -- (4,1.732);
\draw[very thick] (4,0) -- (4.5,0.866);
\draw[line width=1.2mm,postaction={decorate}] (3,1.732) -- (2.5,0.866);
\node at (0,-0.8) {$0$};
\node at (1,-0.8) {$1$};
\node at (2,-0.8) {$2$};
\node at (3,-0.8) {$3$};
\node at (4,-0.8) {$4$};
\node at (5,-0.8) {$5$};
\begin{scope}[xshift=-4cm,yshift=2cm];
\draw[postaction={decorate},very thick] (0,0) -- (1,0);
\draw[postaction={decorate},very thick] (1,0) -- (0.5,0.866);
\draw[postaction={decorate},very thick] (0.5,0.866) -- (0,0);
\node at (0.5,-0.5) {$\Be_1$};
\node at (1.3,0.6) {$\Be_2$};
\node at (-0.3,0.6) {$\Be_3$};
\end{scope}
\end{tikzpicture}
\end{center}
\caption{\footnotesize The structure of vacua and kinks for $\SU(3)$ with rank $k=5$. A kink with charge $\Be_3$ in representation $[\Be_1]$ is shown.}
\label{f6}
\end{figure}

The subtle feature is once again the existence of a hidden RSOS kink structure which is sensitive to $k$ the rank of the spin representation. Generalizing the $\SU(2)$ case above, the vacua are associated to the highest weights of $F$ at level $\leq k$; illustrated in fig.~\ref{f6} for the example of $\SU(3)$ at level $k=5$. The states transforming in the $a^\text{th}$ antisymmetric 
representation $[\Bomega_a]$, where $\Bomega_1=\Be_1+\cdots+\Be_a$, are kinks that interpolate between the vacuum $\Blambda$ and $\Blambda'$ where the selection rule is that $[\Blambda]\overset k\in[\Bomega_a]\times[\Blambda']$, where the $k$ indicates that representations are restricted to those of level $\leq k$.

So putting everything together, we can associate a basis of excitations of mass $m_a$ and rapidity $\theta$ to the creation---or Zamolodchikov---operators ${\cal Z}^{(a)}_{\Balpha;\Blambda,\Blambda'}(\theta)$ where $a$ labels an antisymmetric representation of $\SU(N)$; $\Bv$ is a particular weight of the representation; and $\Blambda\overset k\in[\Bomega_a]\times[\Blambda']$. The $
\SU(N)$ symmetry acts on the $\Bv$ index. A multi-particle state is then generated by acting on the ground state as
\EQ{
{\cal Z}^{(a_n)}_{\Bv_n|{\boldsymbol0},\Blambda_{n-1}}(\theta_n)\cdots  {\cal Z}^{(a_2)}_{\Bv_2|\Blambda_2,\Blambda_1}(\theta_2){\cal Z}^{(a_1)}_{\Bv_1|\Blambda_1,{\boldsymbol0}}(\theta_1)\ket{0}\ .
}

\subsection{Heisenberg XXX$_{k/2}$ spin chain}\label{s5.3}

In this section, we digress to discuss in more detail the relation of our construction with the $XXX_{k/2}$ Heisenberg spin chain with spins in the spin $k/2$ representation for $\SU(2)$ and their generalization to arbitrary groups.

It has been known for long time, that the Heisenberg anti-ferromagnetic $XXX_{k/2}$ chain has gapless regimes in the universality class of WZW models. This connection is described at the phenomenological level in the classic papers by Affleck and Haldane \cite{Affleck:1987ch} and Affleck \cite{Affleck:1985wb}. 

The integrable $XXX_{k/2}$ chain defined by Taktajan \cite{Takt} and Babujian \cite{Babujian:1982ib} is multi-critical lying in the universality class of the WZW with level $k$. This model is related to the light cone model described earlier in the following way. Firstly one defines a chain with the same set of alternating inhomogeneities as before, but with a different Hamiltonian. Following Reshetikhin and Saleur \cite{Reshetikhin:1993wm}, we define
\EQ{
H_\pm&=\frac i{2\Delta}\frac d{dz}\log \Tr_0T(z)\Big|_{z =\pm\nu }\\ &=\frac i{2\Delta}\sum_{n\text{ even/odd}}R_{n+1,n}(\pm2\nu   )^{-1}\dot{R}_{n+1,n}(\pm2\nu   )\\ &\,+\frac i{2\Delta}\sum_{n\text{ odd/even}}R_{n+2,n+1}(\pm2\nu   )^{-1}P_{n+2,n}\dot {R}_{n+2,n}(0)R_{n+2,n+1}(\pm2\nu   )\ ,
} 
where $\dot R=dR/dz$.

The Hamiltonian is then
\EQ{
H=H_++H_-\ .
\label{ham}
}
In contrast to the light cone chain, this Hamiltonian is local on the spin chain.
If $\nu=0$, then, up to a constant,
\EQ{
H=\frac i\Delta\sum_nP_{n+1,n}\dot{R}_{n+1,n}\ ,
}
which, for arbitrary spin $s=k/2$ can be expressed explicitly as
\EQ{
H=\frac1\Delta\sum_{j=0}^{2s}\sum_{\ell=1}^j\frac1\ell\prod_{i=0\atop(\neq j)}^{2s}\frac{\BS_m \cdot\BS_n-x_i}{x_j-x_i}\ .
}
where 
\EQ{
x_j=\frac12j(j+1)-s(s+1)\ .
}
This is precisely the Hamiltonian of the integrable Taktajan-Babujian $XXX_{k/2}$ spin chain.

For the case of spin $s=\frac12$ ($k=1$), and for arbitrary $\nu$, the Hamiltonian \eqref{ham} can be written explicitly as 
\EQ{
H=\frac{2\Delta^{-1}}{1+4\nu^2}\sum_n\Big[\BS_{n+1}\cdot\BS_n+2(-1)^n\nu\BS_{n+1}\cdot\BS_{n}{\boldsymbol\times}\BS_{n-1}
+2\nu^2\BS_{n+1}\cdot\BS_{n-1}\Big]\ .
\label{puy}
}
So at $\nu=0$, this is just the $XXX$ Heisenberg spin chain. On the other hand in the limit of large inhomogeneity $\nu\to\infty$, the spin chain degenerates into 2 decoupled integrable $XXX$ spin chains.

The spin chain is solved by the same ABA method that we described earlier, the only difference is that since the Hamiltonian is different the expressions for the energy of the pseudo particles is changed compared to \eqref{ppe}. On the other hand, the momentum is the same. The new pseudo particle energy is
\EQ{
\varepsilon_0(z)=\frac{2k}\Delta\Big[\frac1{4(z-\nu)^2+k^2}+\frac1{4(z+\nu)^2+k^2}\Big]\  ,
}
c.f.~\eqref{ppe}. The density of $k$ strings in the ground state stays the same since the momentum is the same. For the excitations above the ground state, their energy changes from \eqref{xen} to
\EQ{
\varepsilon(z)=\frac\pi{4\Delta}\big[\sech\pi(z+\nu)+\sech\pi(z-\nu)\big]\ ,\qquad
\wp(z)=\frac2\Delta\tan^{-1}\Big(\frac{\sinh\pi z}{\cosh\pi\nu }\Big)\ .
\label{uil}
}
With $\nu=0$ we get the dispersion relation of the $XXX$ model that we wrote down in \eqref{xxe}.

The interpretation of \eqref{uil} is interesting \cite{Reshetikhin:1993wm}. For fixed $\nu$, the excitations are gapless, however, as $\nu\to\infty$ two Brillouin zones emerge corresponding to $|z|<\nu$ and $|z|>\nu$ centred around $\wp\sim0$ and $\wp\sim\pm\pi/2\Delta$. This reflects the fact that, from \eqref{puy}, we see that in the large $\nu$ limit, the spin chain splits into 2 decoupled $XXX$ spin chains for even and odd sites. So the lattice spacing effectively doubles. In the large $\nu$ limit, states in the central Brillouin zone are gapped and become decoupled from the remaining states. So taking a continuum limit with $\Delta\to0$ and $\nu\to\infty$ with
\EQ{
\frac\pi{2\Delta}e^{-\pi\nu}=mc^2\ ,\qquad c=\frac\pi2\ ,
}
fixed, the dispersion relation \eqref{uil} becomes
\EQ{
&\varepsilon(z)=mc^2\cosh\pi z\ ,\qquad \wp(z)=mc\sinh\pi z\ ,\\
&\text{i.e.}\qquad \varepsilon^2-\wp^2c^2=m^2c^4\ .
}
These are once again relativistic with a speed of light $c=\pi/2$. Note that the decoupling of the massless states is apparent at the level of the S-matrix of the excitations to be discussed in section \ref{s6}. 

The conclusion is that if we use the $XXX$ spin chain itself to act as the lattice regularized theory, instead of the light cone approach, then we get the same massive infra-red excitations but, in addition, also a decoupled massless sector.

\section{The S-matrix}\label{s6}

The next quantity to consider is the S-matrix for the scattering of the excitations above the ground state.  Again, let us focus on the $\SU(2)$ case first. 
The S-matrix of the excitations of the $XXX_{k/2}$ spin chain was determined in \cite{Reshetikhin:1990jn}. It is important that the S-matrix does not depend on the actual choice of the Hamiltonian, so is the same for both the light cone approach and the massive sector of the $XXX_{k/2}$ spin chain. All that changes is the dispersion relation of the excitations.

The S-matrix has a characteristic factored structure of the form
\EQ{
S(\theta)=X(\theta)S^{\SU(2)}(\theta)\otimes S^{\text{RSOS}_k}(\theta)\ ,
\label{ffo}
}
to reflect the explicit $\SU(2)$ symmetry and the hidden kink degrees of freedom. From an S-matrix point of view, the two are separated. In the above, $S^{\SU(2)}(\theta)$ is the $\SU(2)$ rational solution of the Yang-Baxter equation associated to the spin $\frac12$ representation of $\SU(2)$. The second factor $S^{\text{RSOS}_k}(\theta)$ handles the kink structure of the states and is written in the Interaction Round a Face (IRF) form. The final factor $X(\theta)$ is a scalar factor that is needed to ensure that the overall S-matrix satisfies unitarity and crossing symmetry and has the right analytic structure to mesh with the existence of bound states in either the direct or crossed channel.

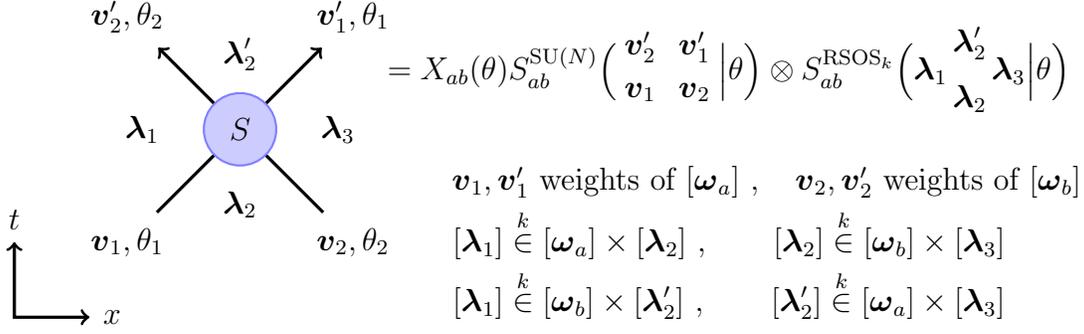
\begin{figure}
\begin{center}
\begin{tikzpicture} [line width=1.5pt,inner sep=2mm,
place/.style={circle,draw=blue!50,fill=blue!20,thick}]
\begin{pgfonlayer}{foreground layer}
\node at (1.5,1.5) [place] (sm) {$S$}; 
\end{pgfonlayer}
\begin{scope}[xshift=-1cm,yshift=0.5cm]
\draw[very thick,<->] (-0.5,-0.5) -- (-0.5,-1.5) -- (0.5,-1.5);
\node at (0.8,-1.5) {$x$};
\node at (-0.5,-0.2) {$t$};
\end{scope}
\node at (0,0) (i1) {$\Bv_1,\theta_1$};
\node at (3,0) (i2) {$\Bv_2,\theta_2$};
\node at (0,3) (i3) {$\Bv'_2,\theta_2$};
\node at (3,3) (i4) {$\Bv_1',\theta_1$};
\node at (1.5,0.5) {$\Blambda_2$};
\node at (0.2,1.5) {$\Blambda_1$};
\node at (2.8,1.5) {$\Blambda_3$};
\node at (1.5,2.5) {$\Blambda'_2$};
\draw[very thick,->] (i1) -- (i4);
\draw[very thick,->] (i2) -- (i3);
\node at (8,2.3) {$=X_{ab}(\theta)S^{\SU(N)}_{ab}\Big(\hspace{-0.1cm}\raisebox{-14pt}{\begin{tikzpicture}[scale=0.6]
\node at (-0.6,0.5) {$\Bv'_2$};
\node at (0.6,0.5) {$\Bv'_1$};
\node at (0.6,-0.5) {$\Bv_2$};
\node at (-0.6,-0.5) {$\Bv_1$};
\end{tikzpicture}}\hspace{-0.1cm}\Big|\theta\Big)\otimes S^{\text{RSOS}_k}_{ab}\Big(\hspace{-0.2cm}\raisebox{-18pt}{\begin{tikzpicture}[scale=0.37]
\node at (0,1) {$\Blambda'_2$};
\node at (0,-1) {$\Blambda_2$};
\node at (1.4,0) {$\Blambda_3$};
\node at (-1.4,0) {$\Blambda_1$};
\end{tikzpicture}}\hspace{-0.2cm}\Big|\theta\Big)$};
\node at (8.5,0.8) {$\Bv_1,\Bv'_1\text{ weights of }[\Bomega_a] \ ,\quad \Bv_2,\Bv'_2\text{ weights of }[\Bomega_b]$};
\node at (8,0.1) {$[\Blambda_1]\overset k\in[\Bomega_a]\times[\Blambda_2]\ ,\qquad [\Blambda_2]\overset k\in[\Bomega_b]\times[\Blambda_3]$};
\node at (8,-0.7) {$[\Blambda_1]\overset k\in[\Bomega_b]\times[\Blambda'_2]\ ,\qquad [\Blambda'_2]\overset k\in[\Bomega_a]\times[\Blambda_3]$};
\end{tikzpicture}
\caption{\small The basic 2-body S-matrix elements. The adjacency condition involve tensor products restricted to the representations of level $\leq k$.}
\label{f7}
\end{center}
\end{figure}

\begin{figure}
\begin{center}
\begin{tikzpicture} [scale=0.5,line width=1.5pt,inner sep=2mm,
place/.style={circle,draw=blue!50,fill=blue!20,thick},proj/.style={circle,draw=red!50,fill=red!20,thick}]
\draw (0,0) -- (2,2);
\draw (4,0) -- (2,2);
\draw[->] (2,4) -- (0,6);
\draw[->] (2,4) -- (4,6);
\node at (2,2) [proj] (p1) {};
\node at (2,4) [proj] (p2) {};
\node at (-0.6,-0.6) (l1) {$a,\theta_1$};
\node at (4.6,-0.6) (l2) {$b,\theta_2$};
\draw[red]  (p2) -- (p1);
\node at (-0.6,6.6) (j1) {$b,\theta_2$};
\node at (4.6,6.6) (j2) {$a,\theta_1$};
\begin{scope}[yshift=-0.5cm]
\node at (11.3,4.6) {\small $\theta=\dfrac{i\pi(a+b)}N$~~~~~~~~~~     for $a+b<N$};
\node at (11.2,2.4) {\small $\theta=\dfrac{i\pi(a+b-N)}N$ ~~~~for $a+b>N$};
\end{scope}
\end{tikzpicture}
\caption{\small The bound states that give rise to poles in the S-matrix at the rapidity difference $\theta=\theta_1-\theta_2$ indicated.}
\label{f8}
\end{center}
\end{figure}
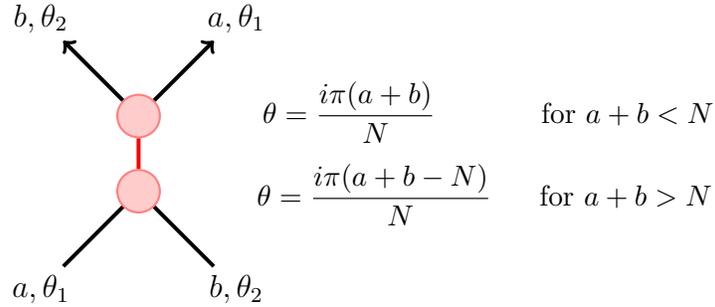

For $\SU(N)$, the S-matrix has the same factored form \eqref{ffo} above. On the creation operators for states, it maps 
\EQ{
Z^{(a)}_{\Bv_1|\Blambda_1,\Blambda_2}(\theta_1)Z^{(b)}_{\Bv_2|\Blambda_2,\Blambda_3}(\theta_2)\longrightarrow
Z^{(b)}_{\Bv'_2|\Blambda_1,\Blambda'_2}(\theta_2)Z^{(a)}_{\Bv'_1|\Blambda'_2,\Blambda_3}(\theta_1)\ .
}
The S-matrix elements are labelled in a way that we summarize in fig.~\ref{f7}. The scalar factor $X_{ab}(\theta)$ has simple poles that correspond to bound states that appear in the direct and crossed channel. The direct channel bound states are shown in fig.~\ref{f8}.

The S-matrix of the lambda model was conjectured, for the case $\SU(2)$, in \cite{Evans:1994hi}. Remarkably it has precisely the same form as the spin chain S-matrix described above. For the $\SU(N)$ generalization, the S-matrix of the current-current deformation of the WZW at level $k$ was conjectured in \cite{Ahn:1990gn}. It is precisely the S-matrix of the $\SU(N)$ spin chain with spins in the rank $k$ symmetric representation \cite{Johannesson:1986ig}. 

It is important to test the S-matrix that we are associating to the lambda model. There are two kinds of test that both rely on a form of the Bethe Ansatz based on the physical excitations and their S-matrix, rather than the pseudo particles of the spin chain. This is the Thermodynamic Bethe Ansatz (TBA). The first way of using it, is to consider the theory at finite temperature $T$. By taking $T\to\infty$ one is probing the UV of the theory, in this case what should be the WZW model. In particular, one can extract the central charge of the UV theory from the S-matrix in a relatively straightforward way yielding
\EQ{
c=\frac{k(N^2-1)}{k+N}\ ,
}
which is precisely the central charge of the $\SU(N)$ WZW model at level $k$.

A second way to test the S-matrix hypothesis, is to use the TBA at $T=0$ but with a background charge, or chemical potential $h$, For $h$ much greater than the mass scale, one can probe the RG flow out of the UV fixed point. The idea is to calculate the free energy density from the S-matrix in this regime and compare it a perturbative calculation from the lambda model which is valid at large $k$. This test is rather sensitive because it should yield the full beta function of the coupling $\lambda$ at large $k$; in other words \eqref{bft}. This test was performed for the $\SU(2)$ case in 
\cite{Evans:1994hi}, although the possibility to extract the beta function for all $\lambda$ in the large $k$ limit was not appreciated. Here, we will generalize the calculation to $\SU(N)$.
 
The idea is to couple the model to a background charge or chemical potential that is very carefully chosen so that the ground state fills up with a single type of particle. The background charge modifies the Hamitlonian to $H\to H-h Q$ and so as $h$ increases beyond a mass threshold, the ground state fills up with particles carrying positive charge. We are interested in the limit of very large $h$ compared with the mass scale. With the carefully chosen charge $Q$, it will be energetically favourable to fill the ground state with particles of the maximal charge and other particles, even if they have positive charge, do not condense in the ground state because they are repelled by the particle with maximal charge.

The free energy density can be calculated on the S-matrix side, by knowing the S-matrix element of the maximally charge particle with itself. This leads to a tractable Weiner-Hopf problem from which the behaviour of the free energy for large $h$ can be calculated. On the Lagrangian side, the free energy density can be calculated in perturbation theory.
A comparison between the two calculations provides a very strong test of the S-matrix and will allow use to extract the beta function at large $k$. In order to complete the picture, the free energy density can also be calculated from the spin chain directly before taking the continuum limit. After taking the continuum limit, we will find that all three approaches give the same result for the free energy density.

\ST{Perturbative calculation}

We now perform the perturbative calculation for the case $F=\SU(N)$.
If we take the action \eqref{tyy} and integrate out the auxiliary field $A_\mu$ what results is a sigma model action for $\CF$ with a WZ term:
\EQ{
S=-\frac k{2\pi}\int d^2x\,\Tr\Big[
\CF ^{-1}\partial_+\CF\big(1+2\lambda\big(1-\lambda\text{Ad}_\CF\big)^{-1}\text{Ad}_\CF\big)\CF ^{-1}\partial_-\CF\Big]+S_\text{WZ}\ .
\label{pp1}
}

The lambda model has a vector symmetry $\CF\to U\CF U^{-1}$ and we can couple the theory to a charge by gauging the symmetry and setting the field to be 
\EQ{
{\cal A}_0=2ih\ ,\qquad {\cal A}_1=0\ ,
}
For the vector symmetry, we effectively replace
\EQ{
\partial_0\CF\rightarrow \partial_0\CF+2ih[Q,\CF]\ ,
}
where $Q$ is the charge, so a constant Hermitian element of the Lie algebra $\mathfrak f$ and $h$ is the chemical potential.

The chemical potential introduces an effective potential (the order $h^2$ terms in the action)
\EQ{
V=-4h^2\Tr\big(\CF^{-1}[Q,\CF]\CF^{-1}[Q,\CF]\big)=-8h^2\Tr\,\big(Q^2-\CF^{-1}Q\CF Q\big)\ .
}
The ground state will be the minimum of this potential. Actually the potential gets modified by the $\lambda$ deformation, but this does not change the conclusion about the ground state since we will be working in the $\lambda\to0$ limit. 
If we expand $V$ about the minimum at $\CF_0$ to linear order, $\CF=\CF_0e^\pi=\CF_0+\CF_0\pi+\cdots$, then
\EQ{
\delta V=-8h^2\Tr\big(\tilde Q[\pi,Q]\big)=-8h^2\Tr\big(\pi[Q,\tilde Q]\big)\ ,
}
where we have defined $\tilde Q=\CF_0 Q\CF_0^{-1}$. 
This vanishes if $[Q,\tilde Q]=0$ which implies that $Q$ and $\tilde Q$ lie in a common Cartan subalgebra. The subgroup of $F$ which fixes a Cartan subalgebra is the lift of the Weyl group which generate permutations in the $N$-dimensional representation.

For simplicity, we will limit ourselves to $N=2n$ even. We will then choose the charge to point along the highest weight of the middle antisymmetric representation  
\EQ{
Q=\Bomega_n\cdot\BH\ ,\qquad\Bomega_n=\Be_1+\Be_2+\cdots+\Be_n\ .
\label{bch}
}
This will ensure that in the S-matrix calculation to come that only one state, namely the one corresponding to the highest weight $\Bomega_n$ will condense in the ground state simplifying the TBA analysis.

The ground state configuration is then (up to the action of the symmetry group)
\EQ{
\CF_0=\left(\begin{array}{c|c}
0 &  1_n\\ \hline
1_n& 0 \\\end{array}\right)\ .
\label{min}
}

The next stage is to work out the action for the fluctuations around the vacuum to quadratic order. We will write
\EQ{
\CF=\CF_0\exp\pi\ ,\qquad 
\pi=\left(\begin{array}{c|c}
\psi_1 &  i\phi_1+\phi_2\\ \hline
i\phi_1^t-\phi_2^t& \psi_2\\
\end{array}\right)\ .
\label{fie}
}
The fields $\psi_i$ do not couple to the charge and so their contribution vanishes when we calculate the shift in the free energy as a function of $h$ relative to $h=0$. Hence the $\psi$ fields can henceforth be ignored.

After some lengthy uninspiring algebra one finds that the quadratic Lagrangian in Euclidean space, after some re-scaling of the fields to bring the two derivative terms into standard form, is
\EQ{
{\cal L}^{(2)}_E&=-\frac{kNh^2}{2\pi}\cdot\frac{1-\lambda}{1+\lambda}+\frac k{4\pi}\Tr\Big[\partial_\mu\phi_i^t\partial_\mu\phi_i-8h\frac{1+\lambda^2}{(1+\lambda)^2}\phi_1^t\partial_1\phi_2\\
&\qquad\qquad +4h^2\Big(\frac{1-\lambda}{1+\lambda}\Big)^4\phi_1^t\phi_1+h^2\phi_2^t\phi_2\Big]\ .
\label{kk1}
}

The goal now is to integrate out the fluctuations $\phi_i$ in the Gaussian approximation. There is a tree level and one loop contribution. We follow  \cite{Evans:1994hi} and use zeta function regularization which is efficient for dealing with the one loop determinant of the non-standard operator we have. The calculation yields the shift in the free energy density,
\EQ{
\delta f(h)=-\frac{Nh^2k}{2\pi}\cdot\frac{1-\lambda}{1+\lambda}-\frac{N^2h^2\lambda^2}{\pi(1+\lambda)^4}\Big(1-\log\frac{8kh^2}{\pi\mu^2}\Big)+\cdots\ ,
}
The terms represented by the ellipsis are higher loop contributions suppressed by further powers of $1/k$, the effective loop counting parameter (or $\hbar$). The parameter $\mu$ is the usual renormalization group scale. The coupling $\lambda$ must run with $\mu$ in order that $\delta f(h)$ is $\mu$ independent at this order in $1/k$. This implies
\EQ{
\frac{Nh^2k}{\pi(1+\lambda)^2}\cdot\mu\frac{d\lambda}{d\mu}+\frac{2N^2h^2\lambda^2}{\pi(1+\lambda)^4}={\cal O}(k^{-1})\ .
}
Hence,
\EQ{
\mu\frac{d\lambda}{d\mu}=-\frac{2N}k\Big(\frac\lambda{1+\lambda}\Big)^2+{\cal O}(k^{-2})\ ,
}
which is none other than the beta function for the $\SU(N)$ case quoted in \eqref{bft}.

Now we integrate the beta function equation and set $\mu$ equal to the physically relevant scale $h$:
\EQ{
\lambda-\frac1\lambda+2\log\lambda=-\frac{2N}k\frac1\xi\ ,\qquad\frac1\xi=\log\frac h{\Lambda}\ ,
}
where $\Lambda$ is the ``$\Lambda$ parameter" of the zeta function regularization scheme. We can solve for the running coupling order by order in $\xi$:
\EQ{
\lambda&=\frac{k\xi}{2N^3}\Big\{N^2-Nk\xi\log\xi+Nk\xi\log\frac{2N}k\\ &\qquad+k^2\Big(1-2\log\frac{2N}k\Big)\xi^2\log\xi+k^2\xi^2\log^2\xi\Big\}+{\cal O}(\xi^3)\ .
}
Plugging into the expression for the shift in the free energy, gives the final result
\EQ{
\delta f(h)&=-\frac{h^2k}{\pi}\Big\{\frac N2-\frac k2\xi+\frac k{4N}\Big[k+N-2k\log\frac{2N}k-N\log\frac{8k}\pi+2k\log\xi\Big]\xi^2\\
&-\frac{k^2}{2N^2}\Big[2k+N-2k\log\frac{2N}k-N\log\frac{8k}\pi+k\log\xi\Big]\xi^3\log\xi+{\cal O}(\xi^3)\Big\}\ .
\label{hij}
}
It is this expression that we will match with the TBA calculation.

\ST{TBA}

From the S-matrix side, the variation of free energy $\delta f(h)$ can be calculated from the TBA equations at $T=0$ coupled to the background charge. These in general, are complicated coupled equations which reflects the fact that many particles can condense in the ground state as the chemical potential is turns on. But with the choice of charge we have made in \eqref{bch}, only one particle contributes to the ground state, namely the one with the biggest charge mass ratio. This is the state with the highest weight in the $\SU(N)$ multiplet $[\Bomega_n]$. It is worth remarking that this state forms no bound states with other asymptotic states and so---intuitively at least---repels other states with positive charge from condensing in the ground state. The same observations were used in \cite{Hollowood:1994np,Evans:1994sv,Evans:1994sy} to calculate the exact mass gaps of a series of integrable models.

Given that only a single particle state condenses in the vacuum, means that TBA equation for its energy satisfies a simple integral equation
\EQ{
\epsilon(\theta)-\frac1{2\pi i}\int_{-B}^B d\theta'\,\epsilon(\theta')\frac d{d\theta}\log S(\theta-\theta')=m\cosh\theta-\frac N2h\ ,
\label{whp}
}
where $S(\theta)$ is the S-matrix element of the highest weight state in the representation $[\Bomega_n]$ with itself and the integration limit $\pm B$ is determined by the condition $\epsilon(\pm B)=0$. Given the solution to \eqref{whp}, the shift in the ground state energy takes the form
\EQ{
\delta f(h)=\frac m{2\pi}\int_{-B}^Bd\theta\,\epsilon(\theta)\cosh\theta\ .
}

The S-matrix kernel in the integral equation \eqref{whp}, can be written in terms of the Fourier transform of a function $R(x)$ as
\EQ{
\frac1{2\pi i}\frac d{d\theta}S(\theta)=\delta(\theta)-\int_0^\infty\frac{dx}\pi\,\cos(x\theta)R(x)\ .
}
For our state, we have
\EQ{
R(x)=\frac{\sinh^2(\pi x/2)}{\sinh(\pi x)\sinh(k\pi x/N)}e^{k\pi x/N}\ .
\label{drr}
}

The solution of the Weiner-Hopf problem \eqref{whp} proceeds by expressing
\EQ{
R(x)=\frac1{G_+(x)G_-(x)}\ ,
}
where $G_\pm(x)$ are analytic in the upper/lower half planes, respectively, and $G_-(x)=G_+(-x)$. The details for solving for $G_\pm(x)$ and extracting the data needed to calculate the shift in the free energy at large $h$ are similar to the $\SU(2)$ case \cite{Evans:1994hi}. One finds
\EQ{
G_+(x)=\sqrt{\frac{4k}N}\frac{\Gamma(1-ix/2)^2}{\Gamma(1-ix)\Gamma(1-ikx/N)}\exp\Big[ibx-\frac{ikx}N\log(-ix)\Big]\ ,
}
where 
\EQ{
b=\frac kN-\frac kN\log\frac kN-\log2\ .
}

The next step is to define a function $\alpha(x)=\exp(2ixB)G_-(x)/G_+(x)$ which has a cut along the positive imaginary axis with a discontinuity that defines $\gamma(\xi)$:
\EQ{
\alpha(i\xi+0^+)-\alpha(i\xi-0^+)=-2ie^{-2\xi B}\gamma(\xi)\ .
}
Hence,
\EQ{
\gamma(\xi)=\exp\Big(-\frac{2k}N\xi\log\xi+2b\xi\Big)\frac{\Gamma(1-\xi/2)^2\Gamma(1+k\xi/N)\Gamma(1+\xi)}{\Gamma(1+\xi/2)^2\Gamma(1-k\xi/N)\Gamma(1-\xi)}\sin(\pi k\xi/N)\ .
}
If we define the expansion
\EQ{
\gamma(\xi)=\pi\exp\Big(-\frac{2k}N\,\xi\log\xi\Big)\sum_{n=1}^\infty d_n\xi^n\ ,
}
then the data that is needed are
\EQ{
&G_+(0)=\sqrt{\frac{4k}{N}}\ ,\qquad
\frac{G_+(0)}{G_+(i)}=\sqrt{\frac{8k}{\pi N}}\ ,\\ &d_1=\frac kN\ ,\qquad d_2=\frac{2k}{N^2}\Big(k\Gamma'(2)-N\log2-k\log\frac kN\Big)\ .
}

The shift in the free energy is a series in powers of $z$ and its logarithm where
\EQ{
\frac1z=2\log\frac hm+2\log\frac{2G_+(0)}{G_+(i)}\ .
\label{frw}
}
The final results to sufficient order to match the perturbative calculation is
\EQ{
\delta f(h)&=-\frac{h^2N^2}{8\pi}G_+(0)^2\Big\{1-2d_1z+\frac{4k}Nd_1z^2\log z\\ &\qquad-2\Big[2d_1-\frac{2k}N\Gamma'(2)d_1-d_1^2+d_2\Big]z^2-\frac{8k^2}{N^2}d_1z^3\log^2z\\ &\qquad+\frac{4k}N\Big[4d_1-\frac{2k}N\Gamma'(3)d_1-2d_1^2+2d_2\Big]z^3\log z+{\cal O}(z^3)\Big\}\ .
\label{sup}
}
Defining 
\EQ{
\frac1\xi=\frac1{2z}+\frac32-\frac12\log\frac{32k}\pi\ ,
}
the expansion of the free energy shift \eqref{sup} is seen to be identical to the perturbative result \eqref{hij}. Then from \eqref{frw} and the above, we extract the exact mass gap of the theory
\EQ{
\Lambda=e^{-3/2}N^{1/2}m\ .
}
Needless to say, the agreement between the perturbative result and the S-matrix TBA calculation provides a very sensitive test of the S-matrix conjecture.

\ST{Spin chain}

The free energy  can also be calculated directly from the spin chain by analysing how the background charge affects the distribution of Bethe $k$-strings in the ground state. 

For the case of $\SU(2)$ the calculation was performed by Faddeev and Reshetikhin \cite{Faddeev:1985qu}. The shift in the ground state energy density is determined by the effective energy of $k$ strings $\varepsilon_k(z)$, 
\EQ{
{\mathscr E}_0(h)={\mathscr E}_0(0)-\frac\pi\Delta\int_{-B}^Bdz\,\Big[\frac1{\cosh\pi(z+\nu)}+\frac1{\cosh\pi(z-\nu)}\Big]\varepsilon_k(z)\ ,
}
where $\varepsilon_k(z)$ satisfies the integral equation
\EQ{
&\varepsilon_k(z)+\int_{-B}^B dw\,J(z-w)\varepsilon_k(w)\\ &\qquad\qquad =-\frac2\Delta\tan^{-1}(e^{-\pi(z+\nu)})+\frac 2\Delta\tan^{-1}(e^{-\pi(z-\nu)})+\frac h2-\frac\pi\Delta\ .
\label{uee}
}
In the above, the Fourier transform of the kernel $J$ is related to $K$ defined in \eqref{dky}:
\EQ{
\tilde J+1=(1+\tilde K)^{-1}=\frac{\tanh(\pi x/2)}{2\sinh(k\pi x/2)}e^{k\pi|x|/2}\ ,
\label{gtr}
}
which is nothing other than the kernel $R$ defined in \eqref{drr} for $N=2$. In \eqref{uee}, the limits of the integrals are defined by the condition $\varepsilon_k(\pm B)=0$.

If we now take the continuum limit as in \eqref{clm}, then 
\EQ{
\frac\pi\Delta\Big[\frac1{\cosh\pi(z+\nu)}+\frac1{\cosh\pi(z-\nu)}\Big]&\longrightarrow m\cosh \pi z\ ,\\
\frac2\Delta\tan^{-1}(e^{-\pi(z+\nu)})-\frac 2\Delta\tan^{-1}(e^{-\pi(z-\nu)})+\frac\pi\Delta&\longrightarrow m\cosh \pi z\ ,
}
and it emerges that the integral equation \eqref{uee} is precisely the integral equation of the TBA \eqref{whp} with the identification $\varepsilon_k(z)=-\epsilon(\theta)$, where $\theta=\pi z$, and 
the shift in the ground state energy density above is exactly the shift in the free energy determined by the TBA earlier:
\EQ{
\delta f(h)={\mathscr E}_0(h)-{\mathscr E}_0(0)\ .
\label{edd}
}
Of course the fact that the two calculations agree is no coincidence because the TBA equations themselves describing the excitations around the ground state can be derived from the spin chain \cite{Reshetikhin:1990jn}. 

The generalization of the analysis to the $\SU(N)$ case proceeds as follows. The kernel $K(z)$ in \eqref{dky}, becomes an $(N-1)\times(N-1)$ matrix kernel with a Fourier transform
\EQ{
K_{ab}(z)=\begin{cases} -1+2\coth(\pi x/N)\sinh(k\pi x/N)e^{-k\pi|x|/N} & a=b\ ,\\
\sech(\pi x/N)\sinh(k\pi x/N)e^{-k\pi|x|/N} & a=b\pm1\ .\end{cases}
}
The equation for the ground state energy generalizing \eqref{uee} involves a matrix kernel $J$ defined as in \eqref{gtr}, with elements
\EQ{
\tilde J_{ab}(x)+\delta_{ab}=\frac{\sinh(\pi(N+1- \text{max}(a,b))x/N)\sinh(\pi \,\text{min}(a,b)x/N)}{\sinh(\pi x)\sinh(k\pi x/N)}e^{k\pi |x|/N}\ .
\label{gtr2}
}
With our choice of background charge only the element with $a=b=N/2$ is relevant and this is seen to be equal to the kernel $R(x)$ in \eqref{drr}. It follows that the change in the ground state energy also matches the change calculated from the S-matrix \eqref{edd}.

\section{Discussion}\label{s7}

In this section, we discuss a series of issues that arise from our investigation of the lambda models and spin chains.

\ST{Recovering non ultra-locality}

To bring the discussion back to the central idea of finding a way around the non ultra-locality problem, we have argued that we can define a discrete version of the theory which is ultra-local and for which the QISM can be applied. The puzzle is that the non ultra-locality is not a pathology of the theory but a feature, so how is it recovered? In the quantum theory of the WZW model, the central term of the Kac-Moody algebra manifests in the behaviour of the two-point function of the current,
\EQ{
\langle \JJ^a_+(z)\JJ^a_+(0)\rangle =\frac {k\delta^{ab}}{z^2}\ .
}
and similarly for $\JJ_-$. For spatial separation $z=ix$.

So from the spin chain point of view, we should expect to see the non ultra-locality re-emerge in the behaviour of these correlators when we choose a scaling limit that focuses in on the WZW model. This involves taking $\nu\to\infty$ and $\Delta\to0$ keeping the mass scale in \eqref{msl} much smaller than the inverse of the distance scales of the correlators. 

Let us focus once again on the $\SU(2)$ case. In Affleck and Haldane's phenomenological theory of spin chains \cite{Affleck:1987ch,Affleck:1985wb}, in the critical regime, the spins at lattice site $n$, $S^a(n)$, are roughly speaking the sum of the holomorphic and anti-holomorphic current plus an alternating term involving the group field $\CF$:
\EQ{
\Delta^{-1}S^a(n)\thicksim \JJ^a(n)+\bar \JJ^a(n)+c(-1)^n\Tr[(\CF-\CF^\dagger)\sigma^a]\ .
}
In particular, this implies that the two-point correlator of say the $z$-component of the spins will behave universally at long distance, compared with the lattice spacing but short compared with the correlation length,  as
\EQ{
\langle S^z(n)S^z(0)\rangle \thicksim-\frac{c_1}{n^2}+(-1)^n\frac{c_2}{n^\eta}\ ,
}
where $\eta=3/(1+2k)$. The $n^{-2}$ term is precisely the manifestation of the non ultra-locality in the classical limit, here emerging through the behaviour of the correlators of the spin chain. The alternating term above has an exponent that is determined by the anomalous dimension of the group field $\CF$. Note that these correlators can be calculated directly within the QISM \cite{Bogolyubov:1986hi,KorepinBook}.

\ST{The PCM}

The principal chiral model is related to the $k\to\infty$ limit of the lambda model. It has been notice that, in this limit, the S-matrix \eqref{ffo} does not give directly the PCM S-matrix \cite{Faddeev:1985qu,Destri:1987ug} which has the form
\EQ{
S^\text{PCM}(\theta)=X(\theta)S^{\SU(2)}(\theta)\otimes S^{\SU(2)}(\theta)\ ,
\label{ffo2}
}
rather it gives an S-matrix that is related to the PCM S-matrix by a vertex-to-IRF transformation \cite{LeClair:1989wy,Bernard:1990cw,Bernard:1990ys} on one of the $\SU(2)$ factors that we call $\SU(2)_R$.\footnote{The other $\SU(2)_L$ symmetry is already manifest in the lambda model S-matrix \eqref{ffo}.} It is natural to interpret this as  an S-matrix manifestation of non abelian T duality \cite{Hoare:2013hhh}. 

The vertex-to-IRF transformation can be thought of as a change of basis in the Hilbert space for the kink factor from the vertex picture where a multi-particle state takes the form
$\ket{\phi_{m_1}(\theta_1)\phi_{m_2}(\theta_2)\cdots\phi_{m_{\mathscr N}}(\theta_{\mathscr N})}$ which transforms in the tensor product representation ${V_{\frac12}}^{\otimes\mathscr N}$ of the spin $\frac12$ representation of $\SU(2)_R$.
This is the ``vertex" basis. On the other hand, the IRF basis corresponds to decomposing the multi-particle states into irreducible representations of $\SU(2)$. In the new basis, the $\mathscr N$-particle states are formed from $\mathscr N$ kinks that interpolate between ``vacua" that correspond to a chain of spins $(j_{\mathscr N},\ldots,j_2,j_1,j_0)$ with $|j_{i}-j_{i-1}|=\frac12$, that label a state of total spin $j_{\mathscr N}$ in the decomposition ${V_{\frac12}}^{\otimes\mathscr N}$:
\EQ{
&\ket{K_{j_{\mathscr N}j_{\mathscr N-1}}(\theta_{\mathscr N})\cdots K_{j_2j_1}(\theta_2)K_{j_1j_0}(\theta_1)}=\sum_{\{m_i=\pm\frac12\}}\left[\ARR{j_{\mathscr N} & \tfrac12 & j_{\mathscr N-1}\\
M_{\mathscr N}& m_{\mathscr N} & M_{\mathscr N-1}}\right]\cdots\\[2pt]
&\qquad\qquad\cdots\left[\ARR{j_2 & \frac12 & j_1\\
M_2 & m_2 & M_1}\right]\left[\ARR{j_1& \tfrac12 &j_0\\
M_1& m_1 & M_0}\right] \ket{\phi_{m_{\mathscr N}}(\theta_{\mathscr N})\cdots\phi_{m_2}(\theta_2)\phi_{m_1}(\theta_1)}\ ,
\label{vir}
}
where $M_{i+1}= M_{i}+m_{i+1}$. In the above, we have used the $3j$ symbols. 

It is clear from \eqref{vir} that the two bases are equivalent if $2j_0>\mathscr N$ (this suggests that as $k\to\infty$ one should also scale $j_0$ to be large as well so that the kink vacuum of the theory lies towards the middle of the allowed set) and there are no restrictions on the left vacuum $j_{\mathscr N}$. For instance the number of states $2^{\mathscr N}$ 
matches in both bases. However, if we are working with periodic boundary conditions then $j_{\mathscr N}=j_0$ and the kink states are restricted to a subset which corresponds to those in the vertex basis with $m_1+m_2=0$. In particular, the number of kinks $\mathscr N$ must be even. It appears that in the $k\to\infty$ limit, we can only recover the states in the PCM if we allow for non-trivial boundary conditions.
This appears related to the observations of Destri and de Vega \cite{Destri:1987ug} who argued from a path integral standpoint that the light-cone lattice formulation in the $k\to\infty$ limit would only give the path integral of the PCM with a projector onto $\SU(N)_R$ singlets, although the point certainly deserves further study.

\ST{Symmetric space lambda models}

More generally one wants to apply the QISM to all the integrable lambda models an, of course, ultimately the string world sheet theory. Let us describe how this should be possible restricting ourselves here to comments on the bosonic lambda models. One begins with the integrable sigma models that have as a target space a symmetric space $F/G$.  This series includes the Type II symmetric spaces of the form $G\times G/G$ as well as the Type I symmetric spaces classified by Cartan (see the book \cite{HEL}). Note that the former sigma models are known to be integrable at the quantum level only if the denominator group $G$ is simple \cite{Abdalla:1982yd}. The quantum integrable theories are then associated to either Type II symmetric spaces, which provide an equivalent way to formulate the PCM for the group $G$, or Type I symmetric spaces of the form
\EQ{
&\SU(N)/\text{SO}(N)\ ,~~~\qquad\SU(2N)/\text{Sp}(N)\ ,\\
&\SU(2)/\text{U}(1)\equiv\mathbb CP^1\ ,\quad
\text{SO}(N)/\text{SO}(N-1)\equiv S^{N-1}\ ,
}
plus a number of possibilities involving the exceptional groups.

The associated lambda models are associated to the $F/F_V$ gauged WZW model,\footnote{In the type II case, each $G$ factor can be associated to a different level.} as in the PCM case, but now the deformation leaves a subgroup $G\subset F$ of the gauge symmetry intact. The lambda model can be thought of as an integrable perturbation of the $F/G_V$ gauged WZW model by a relevant current-current operator that generalizes \eqref{bxx}
\EQ{
S_\lambda[\CF,A_\mu^{(0)}]=kS_\text{WZW}[\CF,A_\mu^{(0)}]+\frac{4\pi\lambda}k \int d^2x\,\Tr\big(\hat\JJ_+^{(1)}\hat\JJ_-^{(1)}\big)+{\cal O}(\lambda^2)\ ,
\label{bxx2}
}
where the superscripts label the $\mathbb Z_2$ decomposition of the Lie algebra of $F$ that results from the fact that $F/G$ is a symmetric space: $\mathfrak f=\mathfrak f^{(0)}+\mathfrak f^{(1)}$ with $\mathfrak f^{(0)}=\mathfrak g$.
 
Let us take the Type II cases first. The lambda models have an S-matrix described by a generalization of \eqref{ffo} where now both S-matrix component factors are of RSOS type:
\EQ{
S(\theta)=X(\theta)S^\text{RSOS}_{k_1}(\theta)\otimes S^\text{RSOS}_{k_2}(\theta)\ .
\label{ffo2}
}
Such an S-matrix is known to describe relevant deformation of the gauged WZW model for $G\times G/G_V$ with levels $(k_1,k_2)$ \cite{Ahn:1990gn}. But this S-matrix also describes the excitations of an RSOS version of the XXZ type spin chain for group $G$ with spins in the rank $k_1$ symmetric representation based on the quantum group $U_q(G)$ with quantum deformation parameter a root of unity $q=\exp[i\pi/(k_1+k_2+c_2(G))]$ \cite{Reshetikhin:1993wm}. It is this spin chain that lies behind the QISM for a Type II symmetric space lambda model.

Now we turn to the Type I symmetric spaces $F/G$. The symmetric space sigma model with target space $F/G$ can be defined by gauging the $G\subset F_R$ symmetry of a PCM based on the group $F$. 
This leads to the lambda model defined above in \eqref{bxx2}. 

However, we can also formulate the Type I symmetric space sigma models in a different way. First of all, a symmetric space $F/G$ has an associated involution $\sigma$ such that $G$ is precisely the subgroup of $F$ fixed under $\sigma$: $\sigma(U)=U$, for $U\in G$. The involution allows us to define the $F/G$ sigma model  as a Type II $F\times F/F$ symmetric space sigma model (or PCM) with a constraint on the field $f=(f_1,f_2)$, $f_i\in F$, of the form $\sigma(f_1,f_2)=(f_2,f_1)$. This alternative formulation allows one to relate the S-matrix for the $F/G$ theory to that of the $F$ PCM \cite{Abdalla:1986xb}. But it also suggests that we can also formulate the associated lambda model as an $F\times F/F_V$ WZW model plus deformation, as for the Type II symmetric space, but with a similar constraint on the field $\sigma(\CF_1,\CF_2)=(\CF_2,\CF_1)$. Note that this requires that the two levels of the WZW are equal $k_1=k_2\equiv k$. 
Following this to its logical conclusion, suggests that the QISM will involve an RSOS version of the $XXY$ type spin chain for group $F$ with $q=\exp[i\pi/(2k+c_2(F))]$ and spins in the symmetric rank $k$ representation along with a projection on the Hilbert space associated to the involution $\sigma$. This suggests that the  S-matrix of the excitations will have the form \eqref{ffo2} with $k_1=k_2\equiv k$ and with a suitable projection on the states, precisely in the way conjectured in \cite{Hollowood:2015dpa}.

\section*{Acknowledgements}

\noindent
CA and DP are supported by STFC studentships.

\noindent
TJH is supported in part by the STFC grant ST/G000506/1. TJH would like to thank Luis Miramontes, David Schmidtt and Benoit Vicedo for discussions.

\end{document}